\documentclass[
preprintnumbers,
amsmath,
prd,nofootinbib,floatfix,11pt,
]{revtex4}

\newcommand\beq{\begin{eqnarray}}
\newcommand\eeq{\end{eqnarray}}
\def\lsim{\mathrel{\rlap{\lower4pt\hbox{$\sim$}}
    \raise1pt\hbox{$<$}}}                
\def\gsim{\mathrel{\rlap{\lower4pt\hbox{$\sim$}}
    \raise1pt\hbox{$>$}}}            

\allowdisplaybreaks
\usepackage{graphicx}
\usepackage{setspace}

\begin{document}
\renewcommand{\theequation}{\arabic{section}.\arabic{equation}}
\renewcommand{\thefigure}{\arabic{section}.\arabic{figure}}
\renewcommand{\thetable}{\arabic{section}.\arabic{table}}

\preprint{NSF-KITP-13-253}

\title{\large \baselineskip=16pt 
Non-universal gaugino masses and semi-natural supersymmetry in view of the Higgs boson discovery}

\author{Stephen P.~Martin}
\affiliation{
\mbox{\it Department of Physics, Northern Illinois University, DeKalb IL 60115}, \\
\mbox{\it Fermi National Accelerator Laboratory, P.O. Box 500, Batavia IL 60510}, {\em and} \\
\mbox{\it Kavli Institute for Theoretical Physics,University of California, Santa Barbara, CA 93106}
}

\begin{abstract}\normalsize \baselineskip=14pt 
I consider models with non-universal gaugino masses at the gauge 
coupling unification scale, taking into account the Higgs boson 
discovery. Viable regions of parameter space are mapped and studied in 
the case of non-universality following from an $F$-term in a linear 
combination of singlet and adjoint representations of $SU(5)$. I 
consider, in particular, ``semi-natural" models that have small $\mu$, 
with gaugino masses dominating the supersymmetry breaking terms at high 
energies. Higgsino-like particles are then much lighter than all other 
superpartners, and the prospects for discovery at the Large Hadron 
Collider can be extremely challenging.
\end{abstract}


\maketitle

\vspace{-0.4cm}

\tableofcontents

\baselineskip=15.4pt

\setcounter{footnote}{1}
\setcounter{figure}{0}
\setcounter{table}{0}

\section{Introduction\label{sec:intro}}
\setcounter{equation}{0}
\setcounter{figure}{0}
\setcounter{table}{0}
\setcounter{footnote}{1}

The explorations of the Large Hadron Collider (LHC) at $\sqrt{s} = 7$ 
and 8 TeV have significantly impacted the parameter space available for 
low-energy supersymmetry as a solution to the hierarchy problem. In the 
once-popular ``minimal supergravity" (or ``constrained minimal 
supersymmetric standard model", CMSSM) scenario, the lower bounds 
\cite{ATLASSUSY1,ATLASSUSY2,ATLASSUSY3,CMSSUSY} on gluino and up-squark and down-squark masses 
are now well over 1 TeV in all cases, and are up to about 1.7 TeV in the 
case that gluino and squark masses are equal. This motivates looking at 
supersymmetric models that instead have non-universal boundary 
conditions on the soft scalar and gaugino masses. Such models can have 
lower detection efficiencies through compression of the superpartner 
mass spectrum. However, the LHC searches still have a significant reach 
\cite{Alves1,Alves2,LeCompte1,LeCompte2} even in the limit of a severely 
compressed superpartner mass spectrum, and for moderate compression, the 
reach is comparable to that for CMSSM models, for fixed gluino and 
squark masses.

The increasing lower bounds on superpartner masses appears to require 
some fine tuning to accommodate
the electroweak scale. (For recent reviews of naturalness in supersymmetry, see 
refs.~\cite{Baer:2012up,Feng:2013pwa,Craig:2013cxa,Baer:2013gva}.) 
In particular, the supersymmetric 
Higgs mass parameter $\mu$ that appears in the 
Higgs potential has to be balanced against the supersymmetry-breaking Higgs mass parameters,
with, including the leading terms from the one-loop effective potential, 
\beq
-\frac{m_Z^2}{2} = |\mu|^2 + m^2_{H_u} + \frac{3 y_t^2}{16 \pi^2} 
\Bigl \{ f(m_{\tilde t_1}^2) + f(m_{\tilde t_2}^2) - 2 f (m_t^2) +
A_t^2 \frac{f(m_{\tilde t_2}^2) - f(m_{\tilde t_1}^2)}{m_{\tilde t_2}^2- m_{\tilde t_1}^2)}
\Bigr \} + \ldots,
\label{eq:tuning}
\eeq 
where terms suppressed by $1/\tan^2\beta$ or by loop factors are omitted,
and $f(x) = x\ln(x/Q^2) - x$, with $Q$ the 
renormalization scale at which all of
the other parameters on the right side
are evaluated as running parameters. Also, $A_t = a_t/y_t$, where 
$a_t (H^+_u \tilde b_L - H^0_u \tilde t_L) \tilde t_R^*$ appears in 
the soft supersymmetry breaking Lagrangian.
 
Increasing bounds on superpartner masses do not imply fine-tuning of $\mu$ 
by itself. This is because $\mu$ is multiplicatively renormalized, and can 
be obtained from dimensionless supersymmetry-preserving couplings (which 
can be small, completely naturally) multiplied by supersymmetry breaking 
parameters, as in either the Kim-Nilles \cite{KimNilles} or 
Giudice-Masiero \cite{GiudiceMasiero} mechanisms or the next-to-minimal 
supersymmetric standard model \cite{NMSSMreview}, for example. However, in 
order to accommodate the known small value of $-m_Z^2/2$, it appears to be 
necessary to ``tune" the remaining terms on the right side of 
eq.~(\ref{eq:tuning}) against $|\mu|^2$. This has lead to the popularity 
of the ideal of ``natural supersymmetry" (see 
e.g.~ref.~\cite{Papucci:2011wy}), in which one argues that therefore 
$|\mu|$ should be not more than a few hundred GeV, and that top squark 
masses (and the gluino mass, which feeds into them through radiative 
corrections) should be not much heavier, perhaps below a TeV or so. 
Because there is no objective measure on parameter space, it is not 
possible to be more precise than this using unambiguous scientific arguments.

This ``natural supersymmetry" parameter space has not yet been 
eliminated by LHC direct searches for the gluino, top squarks, and 
higgsinos, but it is increasingly under tension (see, for example, 
\cite{Evans:2013jna}). Furthermore, the measured Higgs mass is difficult 
to obtain if both top squarks are light. Therefore one might retreat to 
a more limited notion that I will refer to as ``semi-natural 
supersymmetry", in which only $|\mu|^2$ is required to be small (say, 
less than a few hundred GeV). This can be viewed as requiring only one 
tuning; namely, the rest of the right-hand side of 
eq.~(\ref{eq:tuning}). This tuning is simply accepted, as it is 
preferable to the qualitatively more ridiculous tuning associated with 
non-supersymmetric extensions of the Standard Model. From this 
perspective, there is no real problem with the observed Higgs boson 
mass, since one does not require top squarks to be light. The same sort 
of idea has been considered under the name ``Higgsino (LSP) world" in 
\cite{higgsinoworld1}-\cite{Baer:2011ec}, and the phenomenology has been 
studied in depth in \cite{Baer:2011ec} (see also \cite{Han:2013usa})for a realization that is 
qualitatively similar but somewhat different from the present paper. The 
``focus point" scenario \cite{focuspoint1}-\cite{focuspoint6} at large 
scalar masses is another well-known example realizing small $\mu$.

In CMSSM models, the largest contributions to
$m^2_{H_u}$ at the electroweak scale are due to 
the influence of the gluino mass through renormalization group running \cite{KaneKing1,KaneKing2}.
It has long been appreciated and studied
\cite{KaneKing1}-\cite{Cabrera:2013jya}
that if one abandons 
the usual CMSSM boundary condition of equal gaugino masses at the 
scale of apparent unification of gauge couplings, then the little hierarchy problem can be ameliorated. Specifically, this can be accomplished by choosing the gluino mass parameter ($M_3$) to be smaller 
than the wino mass parameter ($M_2$) by a factor of roughly 3 at the scale of the apparent
unification of gauge couplings, $M_U$. 
This leads to smaller values 
of $|\mu|^2$, which is taken here 
as an indirect indicator for semi-natural supersymmetry, as explained above.
There are many ways to 
achieve this, including the possibility \cite{SU5nonuniversal1}-\cite{SO10nonuniversal2}
that the $F$-term that breaks supersymmetry and gives mass to 
the gauginos is not in a pure singlet representation of the global $SU(5)$ 
or larger group that
contains the Standard Model gauge group $SU(3)_C \times SU(2)_L \times U(1)_Y$. 
These models were the 
subject of much study even before the LHC turned on, in part because the supersymmetric 
little hierarchy problem was evident already with the negative Higgs search results of the LEP2 
collider.

Ref.~\cite{Younkin:2012ui} contained a study of this possibility and 
showed the existence of regions of parameter space that feature much less 
extreme cancellation between $|\mu|^2$ and the rest 
of eq.~(\ref{eq:tuning}) than can be found in CMSSM 
for the same gluino and squark mass scales. In particular, regions of 
parameter space were exhibited that obtain small values of $|\mu|$, 
similar to the focus point case and continuously connected to it in 
parameter space, but with the superpartners as light as could be tolerated 
by the direct search limits at the time. However, this paper appeared just 
before the discovery \cite{ATLASHiggs}-\cite{CMScombination} of the Higgs 
boson with a mass near $126$ GeV.  As a consequence, almost all of the 
interesting parameter space chosen in \cite{Younkin:2012ui} is now 
apparently ruled out by the Higgs boson mass. The purpose of the present 
paper is to present a similar study, but now updated to include 
consistency with the Higgs discovery. Here, one should take into account 
the very significant uncertainties in the theoretical prediction of the 
Higgs mass \cite{Ellis:1990nz}-\cite{Feng:2013tvd}. 
As emphasized in \cite{Feng:2013tvd}, it is likely that the 
leading ${\cal O}(\alpha_S^2 y_t^2)$ 3-loop corrections not used in most 
publicly available two-loop programs   
\cite{feynhiggs1}-\cite{Paige:2003mg} and calculations, but appearing in 
\cite{Martin:2007pg,Harlander:2008ju}, and the public 3-loop program H3m 
\cite{Kant:2010tf}, 
will raise the Higgs mass prediction significantly, especially for very 
large top-squark masses. However, there is an effect from the three-loop 
${\cal O}(\alpha_S y_t^4)$ contributions \cite{Martin:2007pg} which appears 
to dilute this effect by perhaps half. In my opinion, this situation 
really just highlights the theoretical uncertainties that are still large 
in the case of one or both top squarks very heavy, despite the great 
efforts that have gone into calculating multi-loop corrections. In the 
following, I will simply use the MSSM model program SuSpect \cite{suspect} 
to translate parameters into physical masses, but then allow the predicted 
value of $M_h$ to fall anywhere in the region from 123 to 128 GeV.

In any study of MSSM parameter space, somewhat arbitrary 
choices must be made in order to keep the presentation finite. (See however \cite{pmssm1,pmssm2,pmssm3}.)
Below, I will choose to consider
only modifications of the CMSSM in which the gaugino mass parameters are non-universal in such a way 
as would follow from $F$-terms in a mixture of a singlet and a ${\bf 24}$ representation of $SU(5)$,
following ref.~\cite{Younkin:2012ui}.
The gaugino masses are parameterized at $M_U$ as 
\beq
M_1 &=& m_{1/2} (c_{24} + s_{24}),
\\
M_2 &=& m_{1/2} (c_{24} + 3 s_{24}),
\\
M_3 &=& m_{1/2} (c_{24} -2 s_{24}),
\eeq
where $m_{1/2}$ is an overall gaugino mass scale, and $c_{24} = \cos\theta_{24}$ and $s_{24} = 
\sin\theta_{24}$, where $\theta_{24}$ is an angle that parameterizes how much of the $F$-term
is in the adjoint representation of $SU(5)$. In particular, $\theta_{24} = 0$ is the usual
CMSSM-like unified gaugino mass case, while $\theta_{24} = \pm\pi/2$ is the case of a pure
${\bf 24}$ of $SU(5)$. Shifting $\theta_{24}$ by $\pi$ corresponds to changing the signs of
all three gaugino masses, which is the same as changing the signs of the scalar cubic couplings
and the $\mu$ term. In the following, I will consider slices of parameter space with fixed $M_3$,
as this corresponds to roughly constant values of the physical gluino mass, often the most important 
figure of merit for present LHC bounds. The sign of $\mu/M_3$ will be taken positive.
For simplicity, I will also follow \cite{Younkin:2012ui}
by keeping the scalar masses universal at $M_U$, parameterized by a variable 
$m_0$ as in the CMSSM. Then, as shown in \cite{Younkin:2012ui}, the parameter space that evades 
direct constraints on superpartners splits into disjoint ``continents" 
(and some small islands) when mapped in the $m_0$ vs. $\theta_{24}$ plane. 
The ``oceans" between the continents (where no viable solutions occur) include the cases
where $\theta_{24}/\pi \approx -0.102$ where $M_2/M_3$ approaches 0,
and $\theta_{24}/\pi \approx 0.148$ where $M_1/M_3$ and $M_2/M_3$ become very large,
and $\theta_{24}/\pi \approx -0.25$ and $0.75$ where $M_1/M_3$ approaches 0.
Unlike \cite{Younkin:2012ui},
I will also consider cases of large scalar cubic couplings, as this facilitates larger $M_h$ 
for fixed values of the other parameters. The most straightforward way of evading the LHC bounds
on superpartner masses is to simply take the gluino mass parameter $M_3$ to be large enough so
that the gluino and up- and down-squarks are heavier than 1.7 TeV.

In the following, I will map out the predicted values of the $\mu$ 
parameter for several parameter space planes, showing how small $\mu$ can 
naturally be obtained outside of the usual focus point region. This 
includes regions of parameter space in which soft supersymmetry breaking 
is dominated by large gaugino masses, with $m_{1/2}^2$ exceeding $m_0^2$ 
and $A_0^2$ 
by orders of magnitude. By doing this, one finds that constraints from 
flavor-violating processes such as $b \rightarrow s \gamma$ and $B_s 
\rightarrow \mu^+\mu^-$ are easily evaded throughout the parameter spaces 
considered below. The contribution to the muon anomalous magnetic moment 
is no worse (and not significantly better) than in the Standard Model. I 
will require that the lightest supersymmetric particle (LSP) is a 
neutralino, and also map out regions of parameter space according to the 
predicted relic density of dark matter (obtained using the program 
micrOmegas \cite{micromegas1,micromegas2,micromegas3,micromegas4}). Here 
it is important to realize that even if the prediction is far off from the 
current value of $\Omega_{\rm DM} h^2= 0.12$ from \cite{WMAP,Planck}, the 
model can still be viable. If the prediction for $\Omega_{\rm DM} h^2$ is 
too low, then axions or something else could be part or all of the dark 
matter. If the prediction for $\Omega_{\rm DM} h^2$ is too high, then the 
lightest neutralino $\tilde N_1$ could decay, either by $R$-parity violation, or 
into some lighter $R$-parity odd particle $\chi$, which would reducing 
$\Omega_{\rm DM} h^2$ by a factor of $M_\chi/M_{\tilde N_1}$ 
\cite{Covi:1999ty}-\cite{Choi:2011yf}. Nevertheless, it is interesting to 
consider the simplest case in which $\tilde N_1$ is assumed to be stable 
with a thermal relic abundance. I will use the term ``allowed neutralino 
dark matter" for the case that the predicted thermal $\Omega_{\rm DM} h^2< 
0.12$, where either $\tilde N_1$ is the dark matter or a subdominant 
component of it, no matter how small. The complementary region where 
$\Omega_{\rm DM} h^2> 0.12$ will also be mapped, and the boundary between 
these two regions is where one can straightforwardly take $\tilde N_1$ to 
be the dominant dark matter component with a thermal relic abundance. 
Here, it will turn out that much of the parameter space is consistent with 
the recent XENON100 \cite{XENON100} and LUX \cite{LUX} direct detection 
constraints.

\section{Models with small stop mixing ($A_0 = 0$)\label{sec:Azero}}
\setcounter{equation}{0}
\setcounter{figure}{0}
\setcounter{table}{0}
\setcounter{footnote}{1}

\begin{figure}[!tp]
 \begin{minipage}[]{0.98\linewidth}
   \includegraphics[width=\linewidth,angle=0]{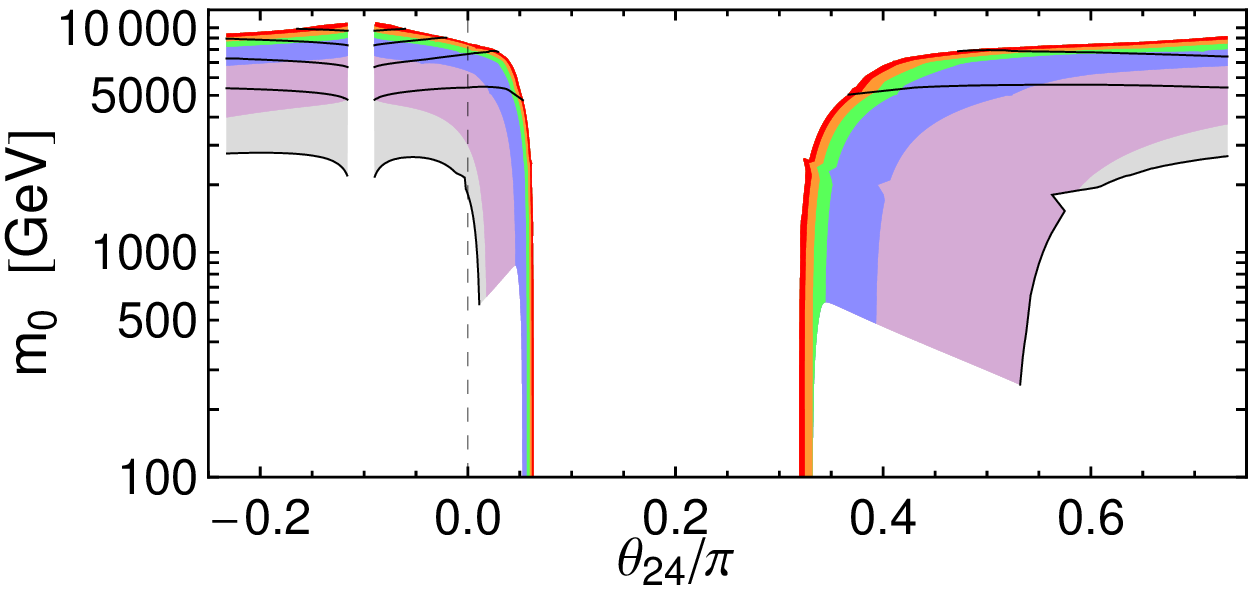}
 \end{minipage}
 \begin{minipage}[]{0.98\linewidth}
   \includegraphics[width=\linewidth,angle=0]{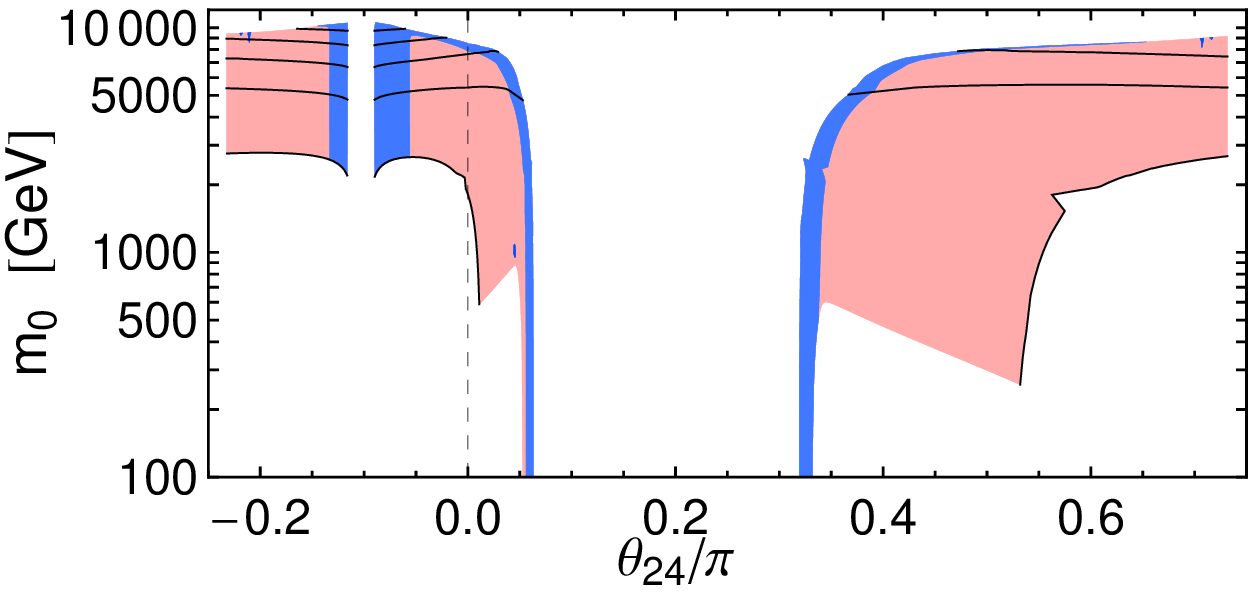}
 \end{minipage}
\caption{\label{fig:tb10_Arat0_2000ALL}
Maps of the $\mu$ parameter (top) and $\Omega_{\rm DM} h^2$ (top), in the 
$m_0$ vs. $\theta_{24}/\pi$ plane, for fixed $\tan\beta = 10$, and $A_0 = 
0$, $M_3 = 2000$ GeV at $M_U$. The vertical dashed line at 
$\theta_{24} = 0$ is the special case of the CMSSM. In each region, the 
lowest black solid line corresponds to $M_h$ = 123 GeV as calculated by 
SuSpect, with each higher black line having $M_h$ increased by 1 GeV. In 
the top figure, the different shaded regions from top to bottom (red, 
orange, green, blue, purple) corresponding to $\mu <$ 500, 750, 1000, 1500 
and 2000 GeV, with the lowest (gray) shaded region for $\mu > 2000$ GeV. 
In the lower figure, $\Omega_{\rm DM} h^2 < 0.12$ is blue and 
$\Omega_{\rm DM} h^2 > 0.12$ is pink. The same color conventions will be used 
throughout this paper.
} 
\end{figure}

First, consider a class of models that is of interest because it includes, 
as a special case, the choice that was used by the LHC 
experimental collaborations to 
constrain CMSSM models with early data at $\sqrt{s} = 7$ and 8 TeV. Take 
$\tan\beta = 10$, $\mu>0$, and $A_0 = 0$ in the usual CMSSM language, and 
allow $m_0$ and $\theta_{24}$ to vary. Maps of the resulting parameter 
space for $\mu$ and for $\Omega_{\rm DM} h^2$ are shown in Figure 
\ref{fig:tb10_Arat0_2000ALL} for the case of $M_3 = 2000$ GeV at $M_U$. 
The allowed parameter space is divided into three 
distinct continents. As explained in \cite{Younkin:2012ui}, the range for 
the plots is chosen to be $-1/4 < \theta_{24}/\pi < 3/4$ in order to avoid 
splitting up the continents unnaturally on the map. The vertical dashed 
line at $\theta_{24} = 0$ represents the special case of CMSSM on this 
plot and similar ones below.

In Figure \ref{fig:tb10_Arat0_2000ALL}a, and similar figures below, the different shaded regions correspond to different
values of $\mu$, according to:
\vspace{-0.3cm}
\beq
\mu < 500\>{\rm GeV}\>\>\> &&{\rm (red)},
\nonumber
\\
500\>{\rm GeV} < \mu < 750\>{\rm GeV}\>\>\> &&{\rm (orange)},
\nonumber
\\
750\>{\rm GeV} < \mu < 1000\>{\rm GeV}\>\>\> &&{\rm (green)},
\nonumber
\\
1000\>{\rm GeV} < \mu < 1500\>{\rm GeV}\>\>\> &&{\rm (blue)},
\nonumber
\\
1500\>{\rm GeV} < \mu < 2000\>{\rm GeV}\>\>\> &&{\rm (purple)},
\nonumber
\\
\mu > 2000\>{\rm GeV} \>\>\> &&{\rm (gray)}.
\nonumber
\eeq
\vspace{-0.8cm}

\noindent The lowest black line on each continent shows where SuSpect predicts $M_h 
= 123$ GeV, with each higher line corresponding to 1 GeV larger for the 
$M_h$ prediction. Recall that the theoretical uncertainties are such that 
a predicted values of 123 GeV may well be consistent with the observed 
Higgs mass. (Boundaries of the shaded region that are not black solid 
lines correspond to the requirements of a neutral LSP, no charged 
superpartner less than 100 GeV accessible to LEP searches, 
or correct electroweak symmetry breaking.) 
An important feature of this parameter space is that as $\theta_{24}$ 
increases from 0, the region compatible with the observed Higgs mass 
increases, extending down to lower values of the scalar masses as 
parameterized here by $m_0$. This is because of the relatively larger 
top-squark mixing.

In each of the continents in Figure \ref{fig:tb10_Arat0_2000ALL}b, the 
region with $\Omega_{\rm DM} h^2 < 0.12$ is shown in blue, and the region 
with $\Omega_{\rm DM} h^2 > 0.12$ is in pink. The boundary between these 
two shaded regions agrees with the cosmological results from the WMAP 
\cite{WMAP} and Planck \cite{Planck} experiments. Several regions can be 
seen in Figure \ref{fig:tb10_Arat0_2000ALL}b to have allowed neutralino 
dark matter. In the CMSSM case ($\theta_{24}=0$), the focus point region 
\cite{focuspoint1}-\cite{focuspoint6} occurs at $m_0 \approx 7900$ GeV. As 
one moves to larger values of $\theta_{24}$, this region with allowed 
neutralino dark matter and small $\mu$ moves to much lower values of 
$m_0$, until for $\theta_{24}/\pi$ near $0.056$ it extends down to very 
small $m_0$. The plot is cut off at $m_0 = 100$ GeV for artistic reasons, 
but in fact the allowed region even extends down to negative $m_0^2$. 
Unlike the situation in the CMSSM, this is possible because the LSP is a 
higgsino-like neutralino here, rather than a charged slepton, 
due to the small value of 
$\mu$, as can be seen in Figure \ref{fig:tb10_Arat0_2000ALL}a. It is 
interesting that the parameter space thus allows gaugino mass domination 
of the soft terms, 
consistent with the constraints of proper electroweak 
symmetry breaking, $M_h$, and allowed neutralino dark matter, and
providing a  solution of the supersymmetric flavor problem
similar to ``no-scale" \cite{noscale1,noscale2,noscale3} or 
``gaugino mediated" 
\cite{gauginomediation1,gauginomediation2,gauginomediation3} models. 

On the other side of the CMSSM-like continent in Figure 
\ref{fig:tb10_Arat0_2000ALL}, with $\theta_{24} < 0$, the focus point 
region occurs at larger values of $m_0$. On the left 
side of this continent, on a vertical line with $\theta_{24}/\pi \approx 
-0.056$, there is sufficient wino mixing in the LSP to give efficient dark 
matter co-annihilation \cite{winocoannihilation}-\cite{Ibe:2013pua}. This 
corresponds to a mostly bino LSP with winos that are only 30 GeV or so 
heavier. In the corner of parameter space with large $m_0 \sim 10$ 
TeV and $\theta_{24}/\pi \approx -0.056$, the focus point and small $M_2$ 
regions merge in a realization of the ``well-tempered neutralino" 
\cite{ArkaniHamed:2006mb} with $\mu$ and $M_2$ both comparable to $M_1$, 
so that the LSP is a mixture of bino, wino, and higgsino in the right 
proportions to allow efficient annihilation of dark matter in the early 
universe.

In the right-most continent in Figure \ref{fig:tb10_Arat0_2000ALL}, 
extending from $0.33 < \theta_{24}/\pi < 0.72$, there is likewise a 
focus-point-like region with large $m_0$. This region is again continuously 
connected to a small $\mu$ region, here along the continent's left edge at 
$\theta_{24}/\pi \approx 0.33$. This small-$\mu$ region once again extends 
down to negligibly small $m_0$, with gaugino mass dominance giving a 
solution to the supersymmetric flavor problem, and small values of $\mu$ 
providing semi-natural supersymmetry.

On the left-most continent, with $-0.232 < \theta_{24}/\pi < -0.116$, 
there is still a very thin focus-point region with small $\mu$ at large 
$m_0$, but this does not connect to any regions with small $m_0$. In both 
the left-most and right-most continents, there are small regions at very 
large $m_0$ in which allowed neutralino dark matter is achieved through 
near-resonant bino-like LSP annihilation through $s$-channel $Z$ and $h$ 
exchange \cite{Drees}. These small regions, with $m_{\tilde N_1}$ near 
$M_Z/2$ and $M_h/2$ respectively, are centered near $\theta_{24}/\pi = 
-0.219$, $0.716$ and near $-0.211$, $0.707$ respectively. In these models, 
the only superpartners that are kinematically accessible to the LHC 
besides the bino-like LSP are the higgsino-like $\tilde N_2, \tilde N_3$ 
and $\tilde C_1$ with typical masses of several hundred GeV.

The existence of the separate continents as seen in Figure 
\ref{fig:tb10_Arat0_2000ALL} is a generic feature of the parameter space 
of models described by a mixture of $SU(5)$ singlet and adjoint $F$-term 
gaugino masses. However, for simplicity, in the remainder of this paper I 
will restrict attention to the CMSSM-like continent that is continuously 
connected to $\theta_{24}=0$.

\begin{figure}[!tp]
 \begin{minipage}[]{0.494\linewidth}
   \includegraphics[width=\linewidth,angle=0]{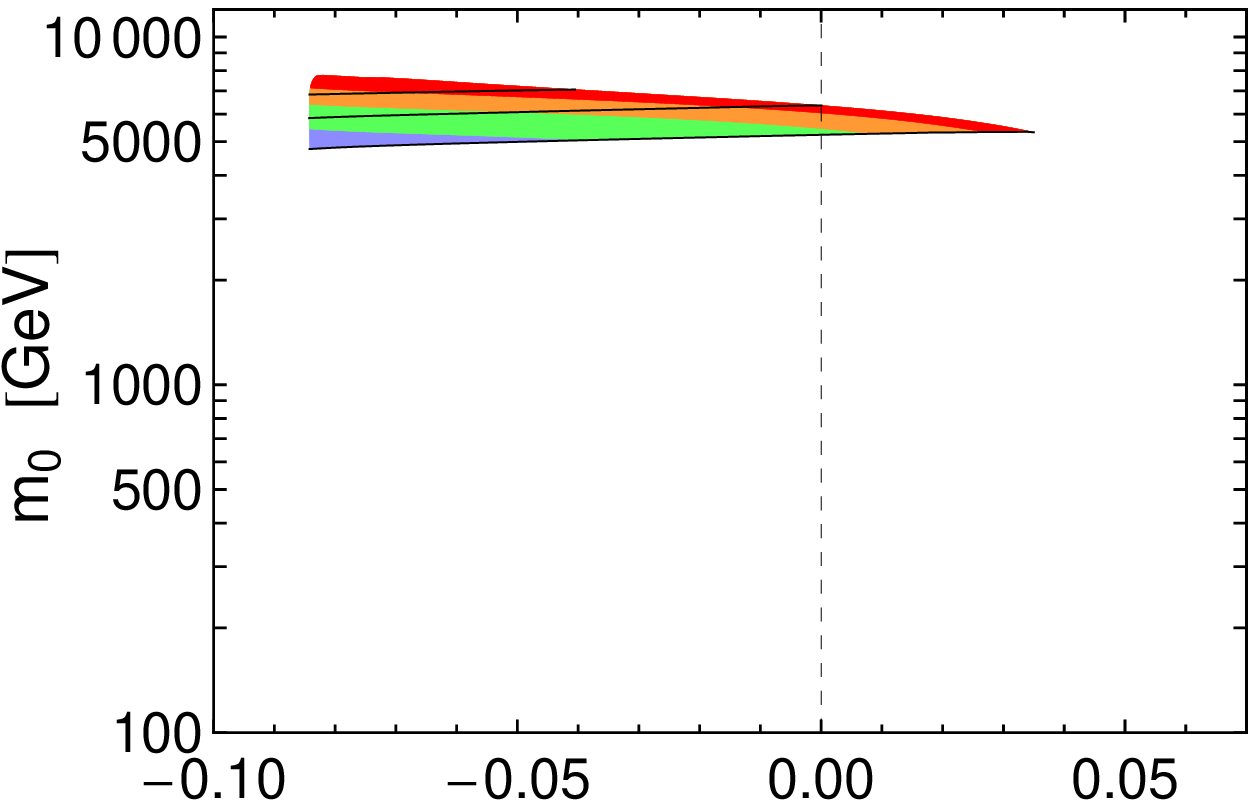}
 \end{minipage}
 \begin{minipage}[]{0.494\linewidth}
   \begin{flushright}
   \includegraphics[width=\linewidth,angle=0]{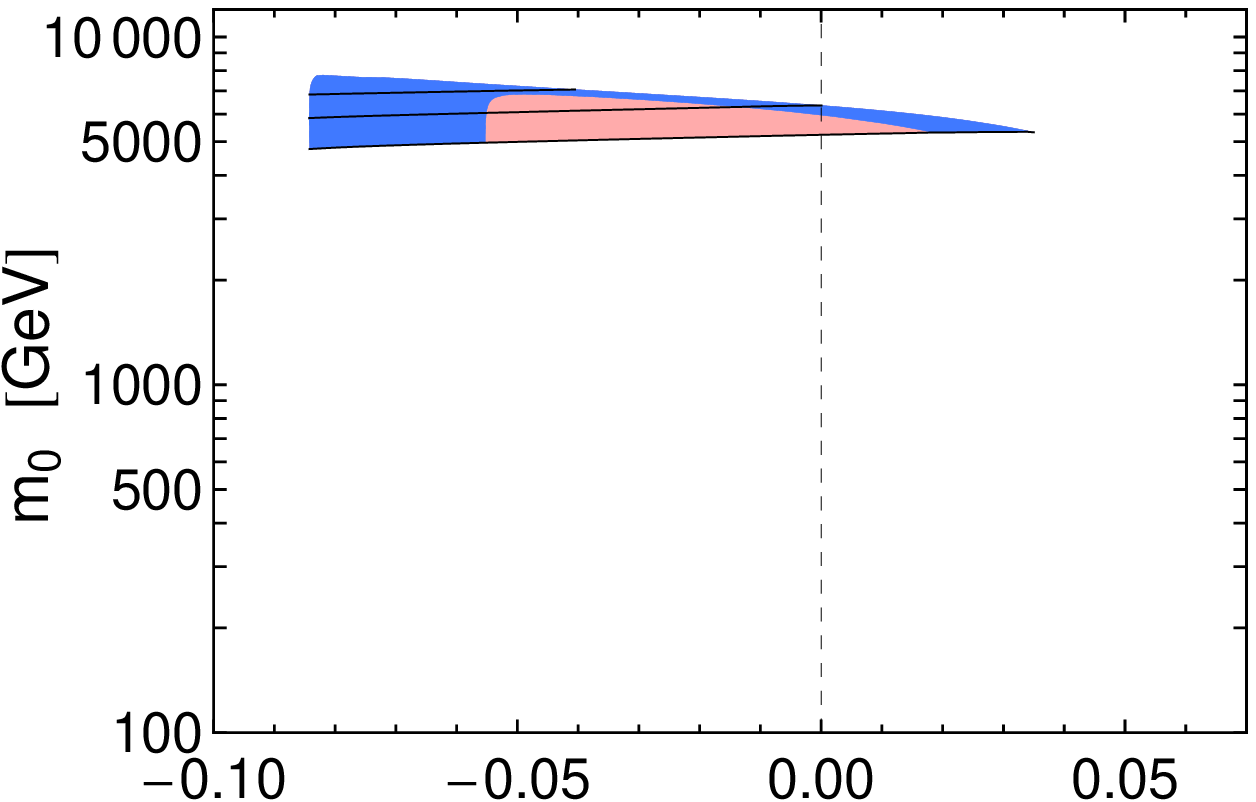}
   \end{flushright}
 \end{minipage}
 \begin{minipage}[]{0.494\linewidth}
   \includegraphics[width=\linewidth,angle=0]{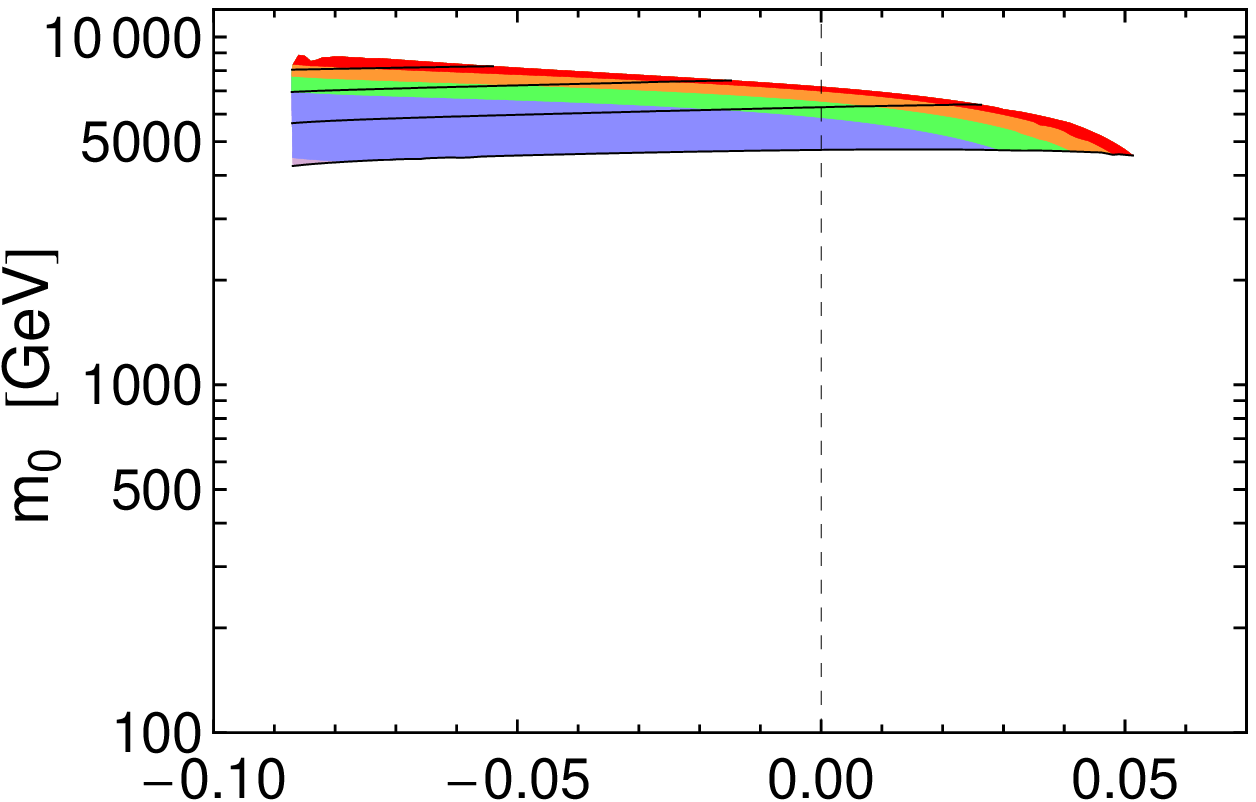}
 \end{minipage}
 \begin{minipage}[]{0.494\linewidth}
   \begin{flushright}
   \includegraphics[width=\linewidth,angle=0]{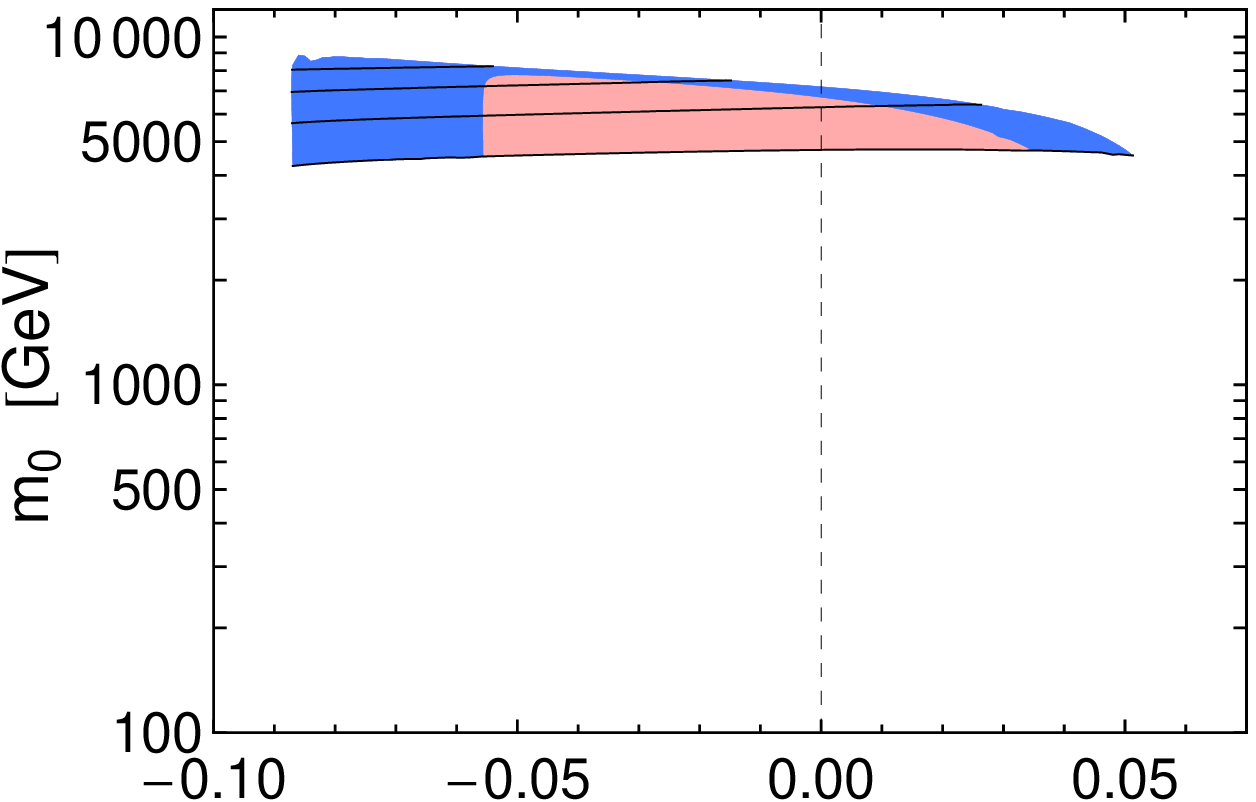}
   \end{flushright}
 \end{minipage}
 \begin{minipage}[]{0.494\linewidth}
   \includegraphics[width=\linewidth,angle=0]{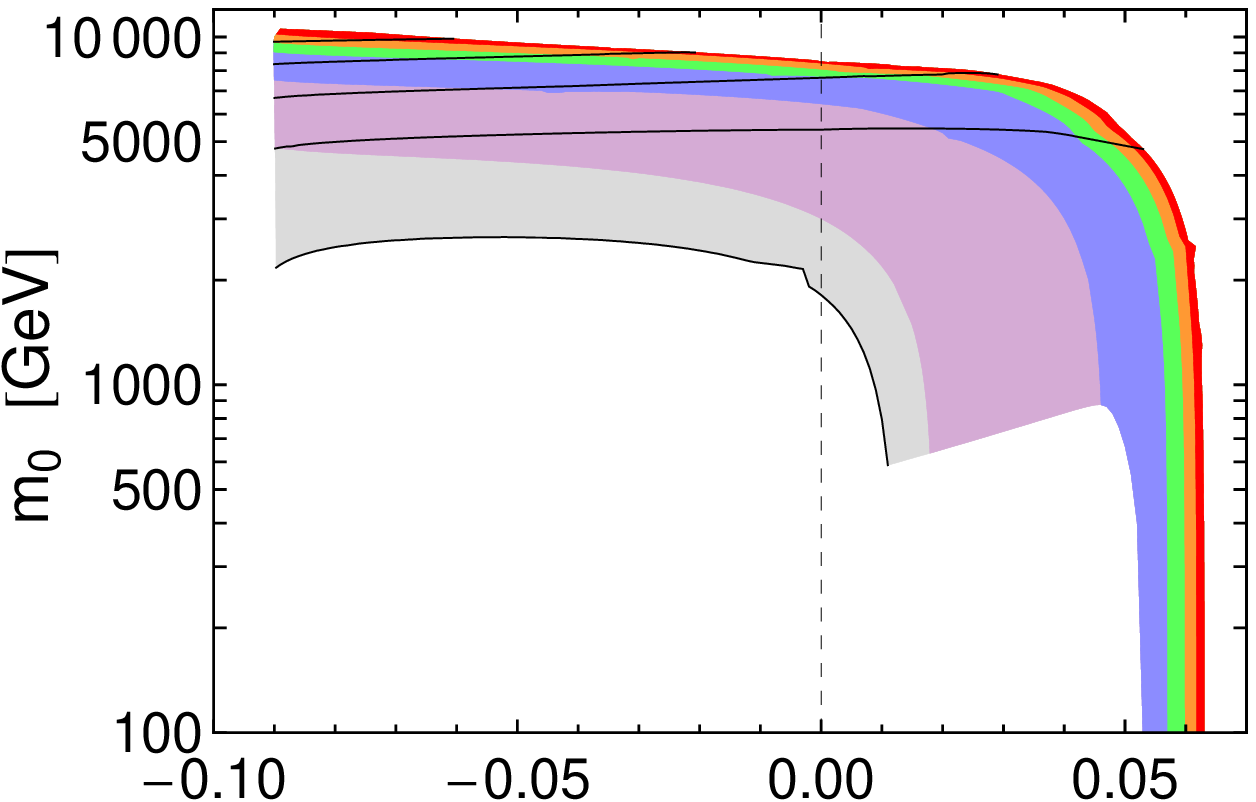}
 \end{minipage}
 \begin{minipage}[]{0.494\linewidth}
   \begin{flushright}
   \includegraphics[width=\linewidth,angle=0]{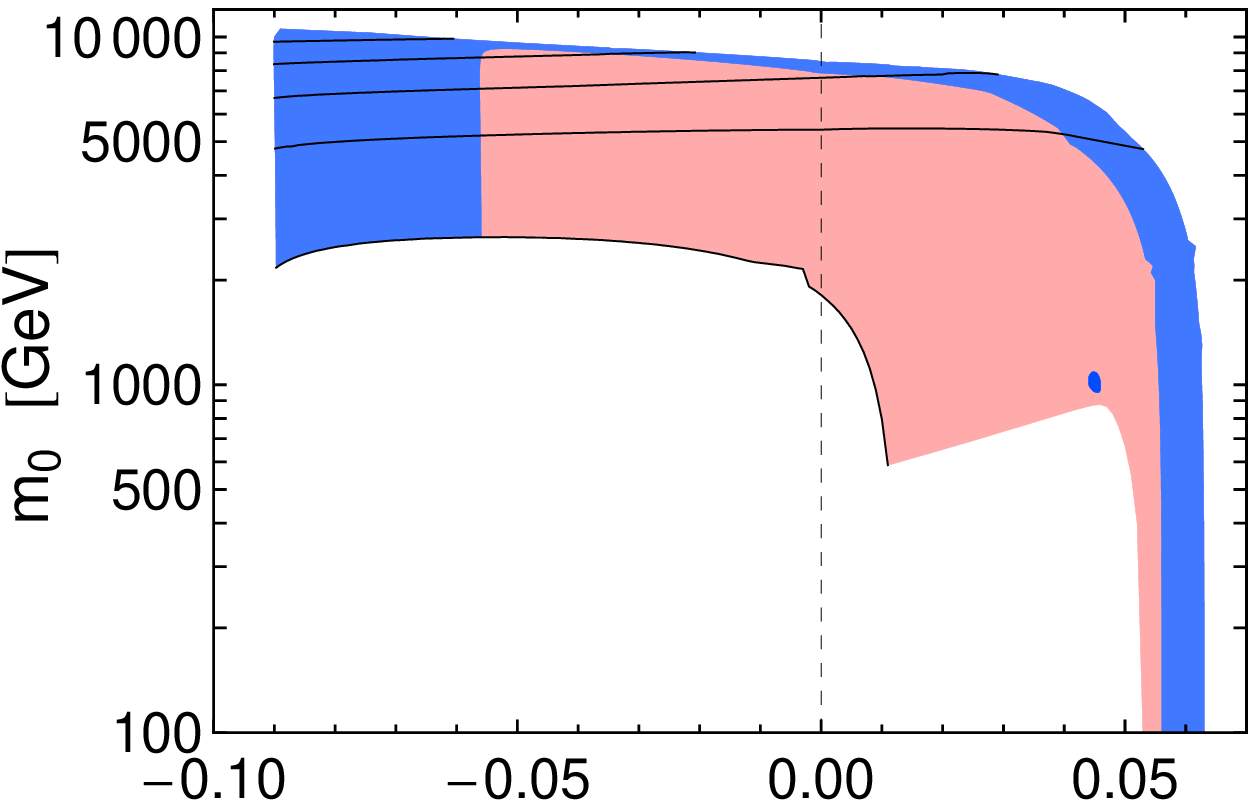}
   \end{flushright}
 \end{minipage}
 \begin{minipage}[]{0.494\linewidth}
   \includegraphics[width=\linewidth,angle=0]{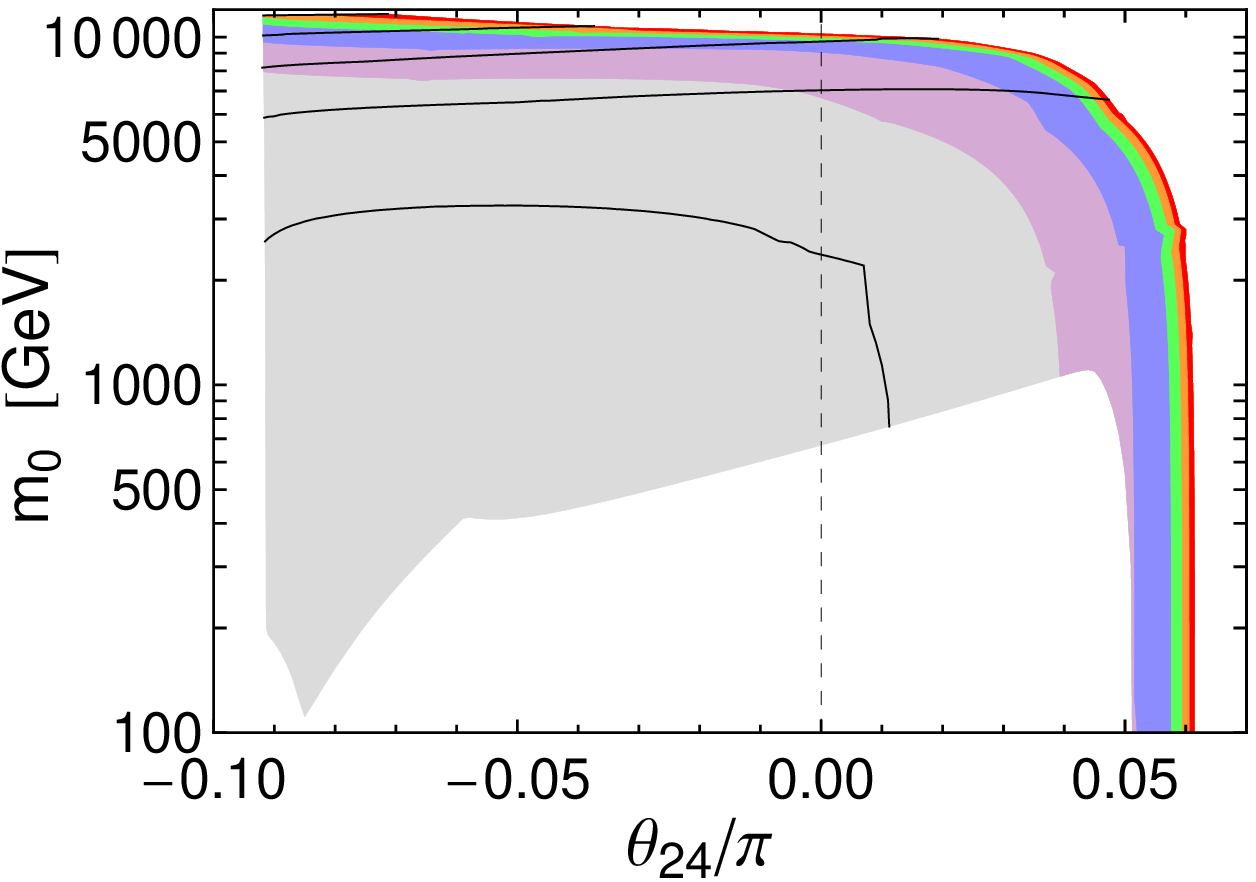}
 \end{minipage}
 \begin{minipage}[]{0.494\linewidth}
   \begin{flushright}
   \includegraphics[width=\linewidth,angle=0]{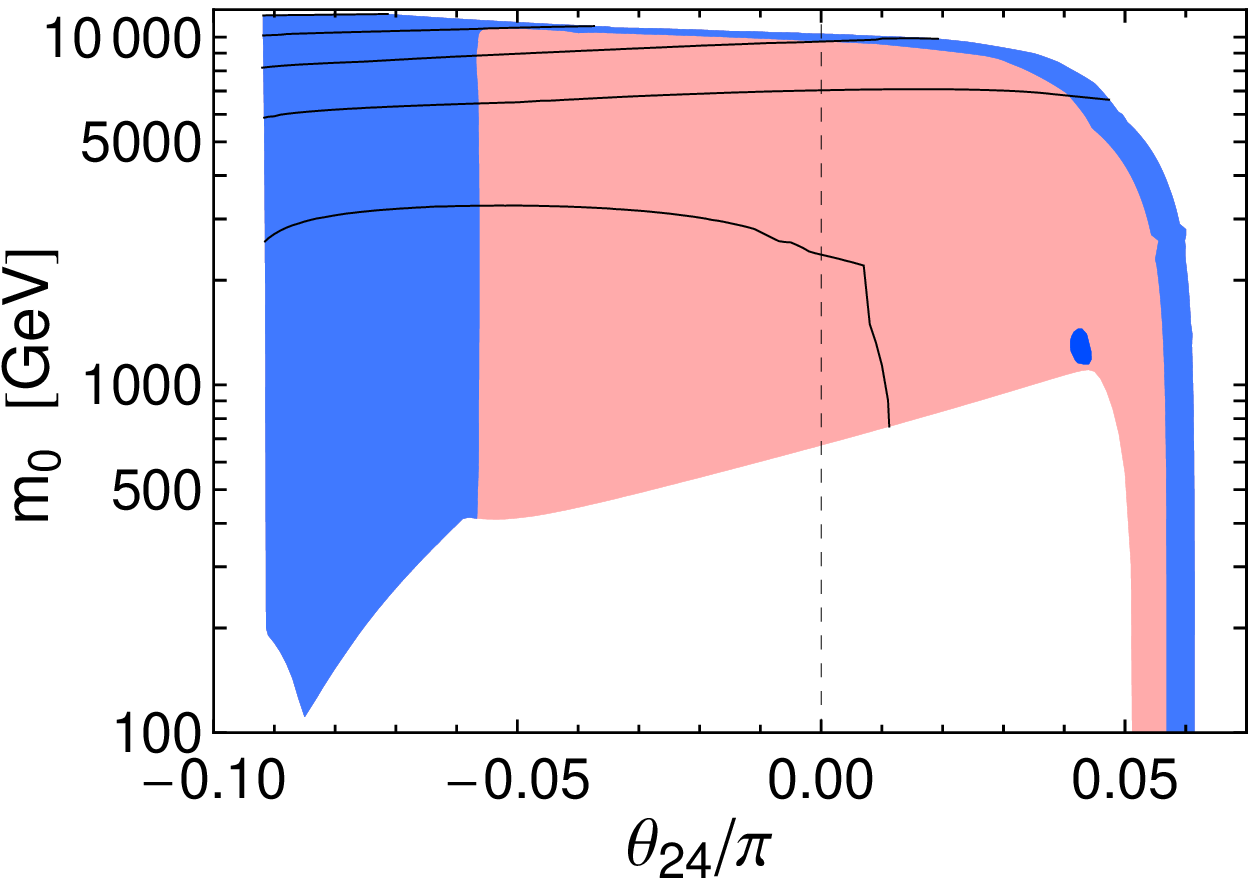}
   \end{flushright}
 \end{minipage}
\caption{\label{fig:tb10_Arat0}
Maps of 
the $\mu$ parameter 
(left) and 
$\Omega_{\rm DM} h^2$ 
(right), as 
in Figure \ref{fig:tb10_Arat0_2000ALL}, but with fixed $M_3 = 1200$, 1500, 
2000, and 2500 GeV (from top to bottom) at $M_U$. In each case, $\tan\beta 
= 10$, and $A_0=0$ at $M_U$. In the case of $M_3 = 2500$ GeV, the lowest 
solid black line corresponds to a SuSpect prediction of $M_h= 124$ GeV, 
while in the other cases it is $M_h = 123$ GeV.}
\end{figure}

The maps of the $\mu$ parameter and allowed neutralino dark matter, for 
models consistent with $M_h$ and other constraints, are shown in Figure 
\ref{fig:tb10_Arat0} for $M_3 = 1200$, 1500, 2000, and 2500 GeV. In each 
case, $A_0 = 0$, and $\tan\beta=10$, with $\mu>0$, just as in Figure 
\ref{fig:tb10_Arat0_2000ALL}. The $M_3 = 1200$ case was chosen as this is 
one of the lowest values for which $M_h$ is consistent with 
observation in an appreciable region of parameter space, although one must 
have $m_0 \gsim 5000$ GeV. The reason that significantly lower $M_3$ will 
not work with these parameters is that to try to accommodate $M_h$ one must 
increase the scalar masses (through $m_0$) 
so much that electroweak symmetry breaking does not work 
(the solution for $|\mu|^2$ from the effective potential becomes 
negative).

For both $M_3=1200$ and 1500 GeV,
the resulting allowed neutralino dark matter region in Figure \ref{fig:tb10_Arat0} consists of a focus-point
region with small $\mu$ that is continuously connected to a region on the left (with 
$\theta_{24}/\pi \lsim -0.055$) were $M_2$ is smaller, giving a 
bino-like neutralino LSP with significant wino content.
For $M_3 = 1500$ GeV, this region extends to slightly smaller values of $m_0$, but one must still have
$m_0$ larger than 4000 GeV. 

For larger $M_3$, the region consistent with $M_h$ in Figure 
\ref{fig:tb10_Arat0} increases dramatically. As noted above, for $M_3 = 
2000$ GeV and $M_3 = 2500$ GeV, the region of allowed neutralino dark 
matter then extends to very small values of $m_0$ on the right side of the 
continent, with $\theta_{24}/\pi \gsim 0.05$. (Note that the $M_3 = 2000$ 
GeV case is just a close-up of the full map shown in Figure 
\ref{fig:tb10_Arat0_2000ALL}.) In particular, for $\theta_{24} > 0$, the 
viable region where $M_h$ is large enough is seen to include much lower 
values of $m_0$ than in CMSSM, and even extends down to $m_0^2 < 0$. This also 
coincides with a region of allowed neutralino dark matter with small 
$\mu$. This region is continuously connected to the focus point region. 
The region is particularly attractive because the gaugino mass dominance 
provides a natural solution to the supersymmetric flavor problem; flavor 
violation is suppressed by the large ratio of gaugino masses to scalar 
masses yielding a nearly flavor-bind sfermion sector. Here the scalar 
squared masses are parameterized by the small flavor-preserving $m_0^2$, 
but in general they could be any similarly 
small scalar squared masses with 
arbitrary flavor structure without significantly affecting the results. In 
both the $M_3=2000$ and 2500 GeV plots, the lower boundary of the shaded 
allowed region is set by the imposed requirement that the LSP not be a 
charged stau. (If $R$-parity is violated, this requirement could be relaxed.)
Note that because of the requirement of accommodating $M_h$, the 
physical gluino mass has to be of order at least about 3 TeV, so that LHC 
discovery prospects may have to hinge on discovery of the lighter bino- 
and higgsino-like neutralino and chargino states instead. This will 
require very high luminosity because of the low cross-section, and may be 
quite problematic.

In the $M_3=2000$ and 2500 GeV plots of Figure \ref{fig:tb10_Arat0}, there 
is also a small island of allowed neutralino dark matter visible centered 
near $(\theta_{24}/\pi, m_0) = (0.043, 1300$ GeV) in the latter plot, 
on the shores of which 
$\Omega_{\rm DM} h^2 = 0.12$. Here the LSP is mostly bino-like but with
a significant mixing with somewhat heavier higgsinos, and efficient annihilation of dark 
matter in the early universe crucially relies on the near-resonant 
co-annihilation process through a charged Higgs scalar boson: $\tilde N_1 
\tilde C_1^\pm \rightarrow H^\pm \rightarrow tb$, with an important 
role played also by neutral Higgs-mediated co-annihilations 
$\tilde N_1 \tilde N_{2,3} \rightarrow b\overline b$ mediated by neutral Higgs bosons
$A^0$ and 
$H^0$. The higgsino-like $\tilde N_2, \tilde C_1$ can be up to about 225 GeV heavier than the LSP here.

\section{Models with moderate stop mixing ($A_0 = -m_0$)\label{sec:Aminusone}}
\setcounter{equation}{0}
\setcounter{figure}{0}
\setcounter{table}{0}
\setcounter{footnote}{1}

\begin{figure}[!tp]
 \begin{minipage}[]{0.494\linewidth}
   \includegraphics[width=\linewidth,angle=0]{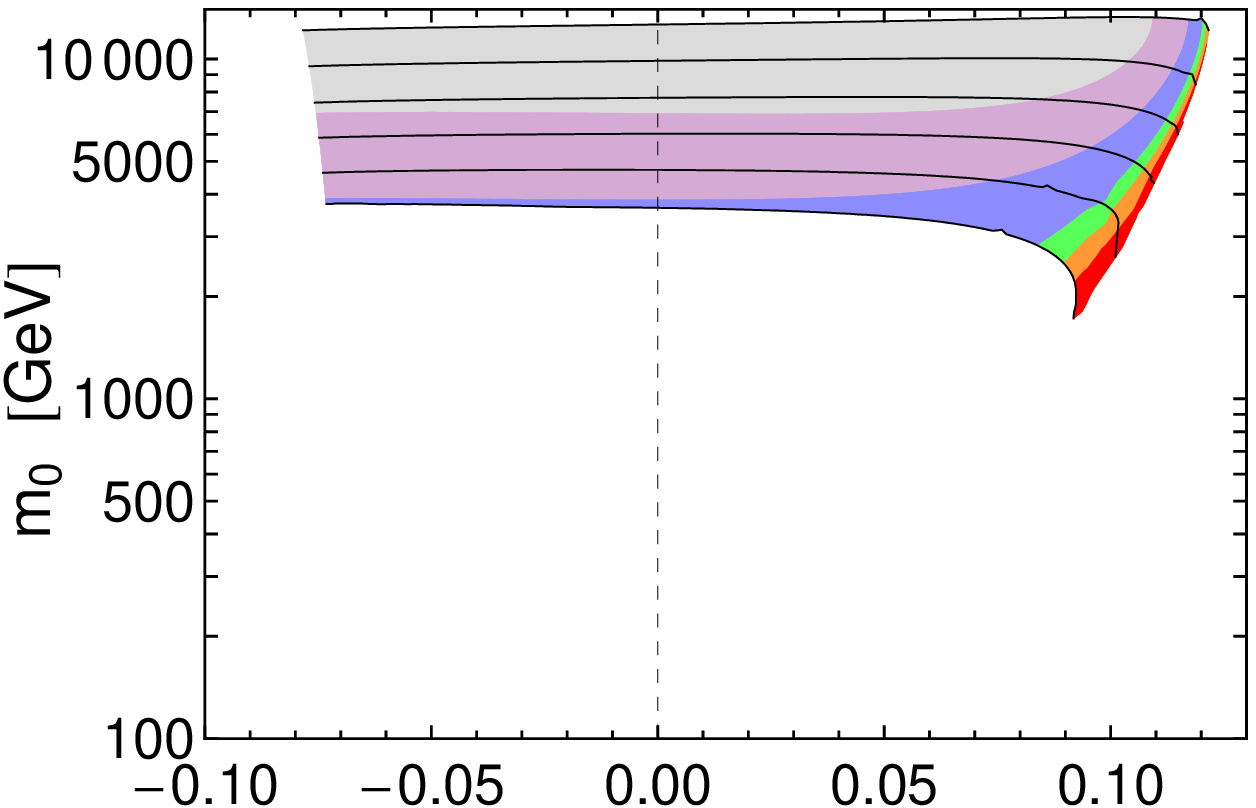}
 \end{minipage}
 \begin{minipage}[]{0.494\linewidth}
   \begin{flushright}
   \includegraphics[width=\linewidth,angle=0]{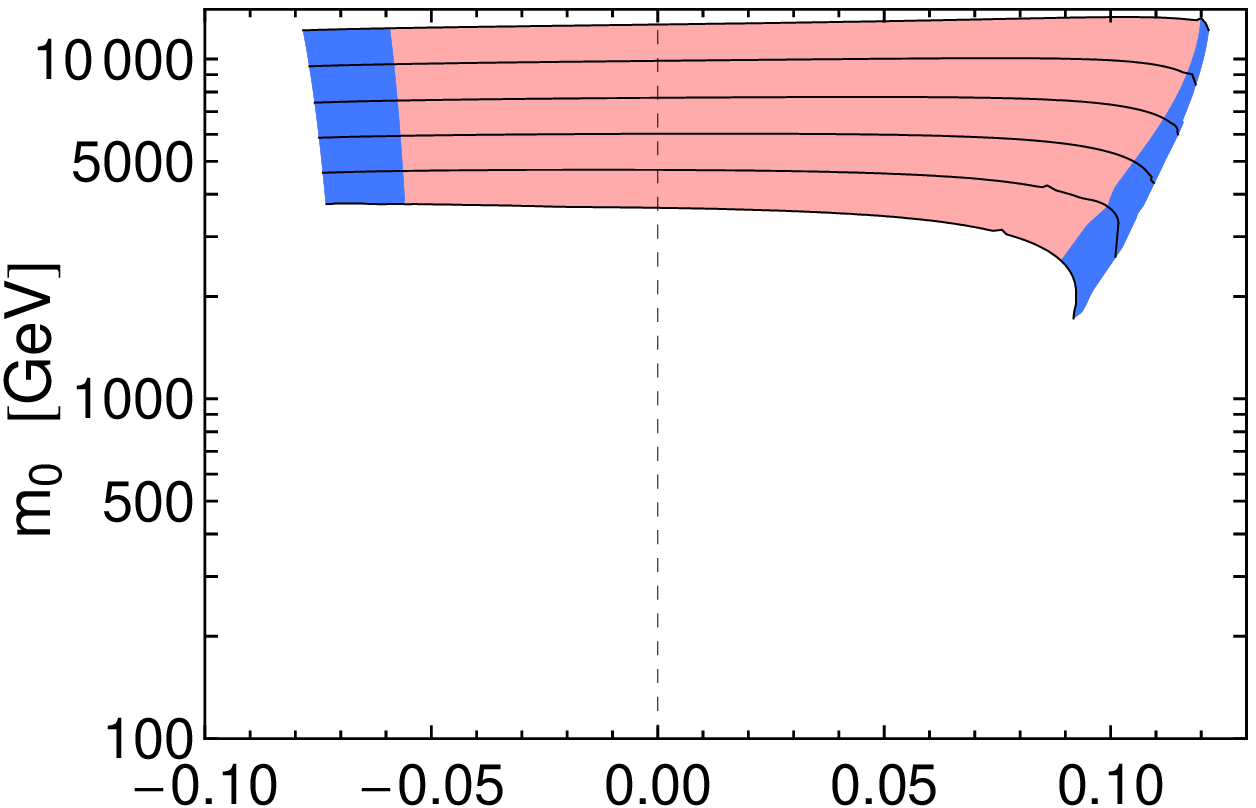}
   \end{flushright}
 \end{minipage}
 \begin{minipage}[]{0.494\linewidth}
   \includegraphics[width=\linewidth,angle=0]{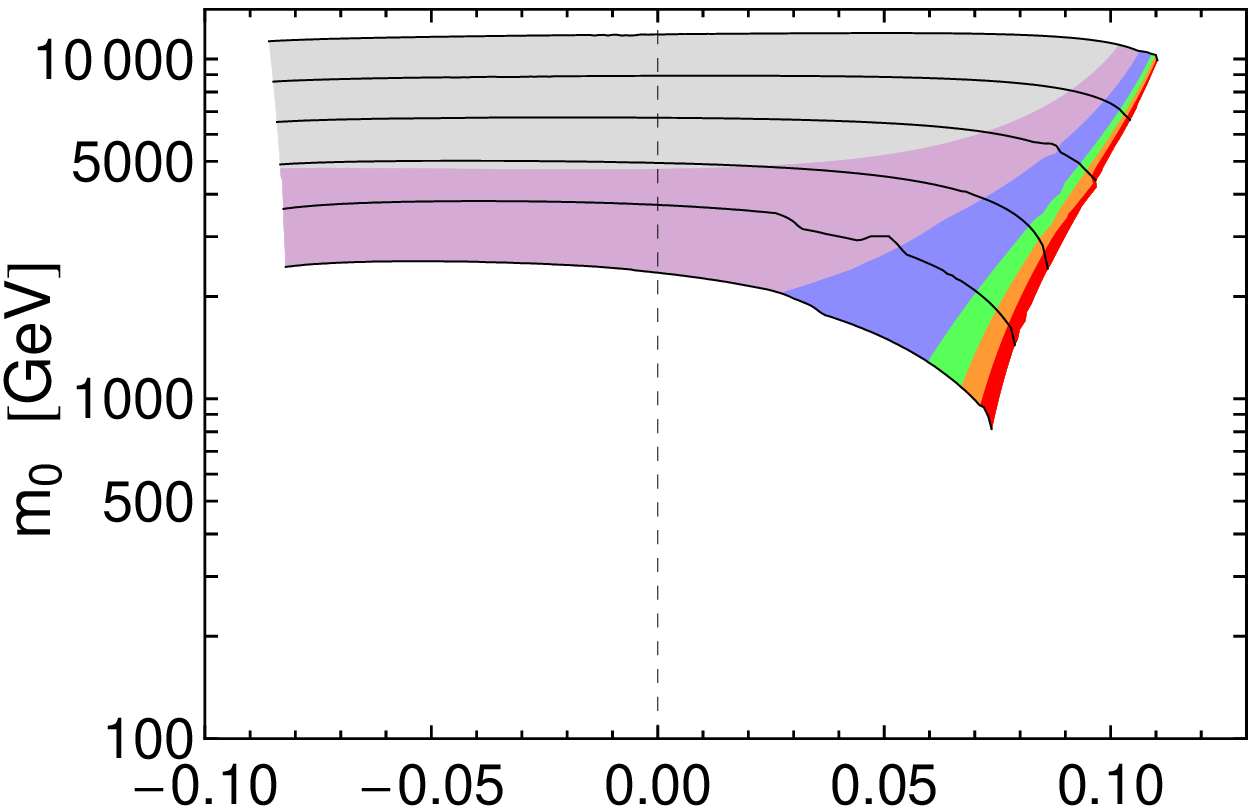}
 \end{minipage}
 \begin{minipage}[]{0.494\linewidth}
   \begin{flushright}
   \includegraphics[width=\linewidth,angle=0]{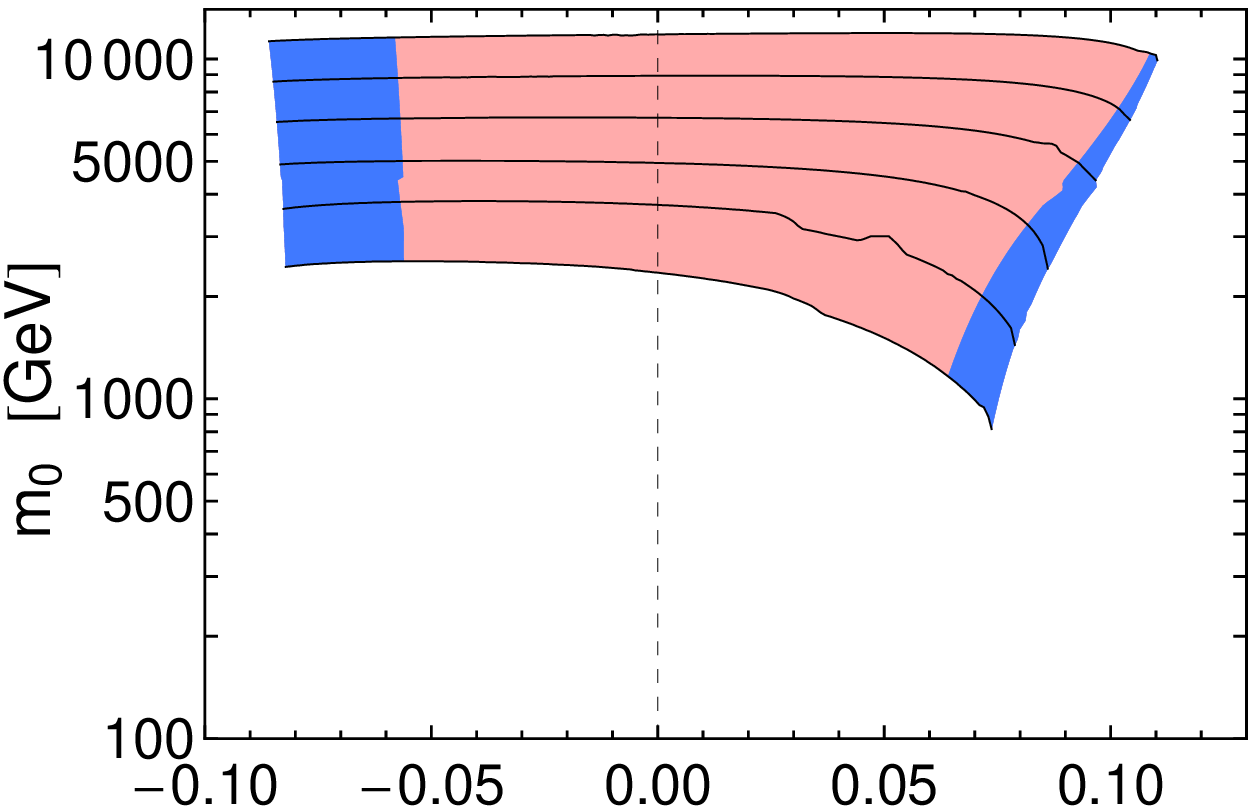}
   \end{flushright}
 \end{minipage}
 \begin{minipage}[]{0.494\linewidth}
   \includegraphics[width=\linewidth,angle=0]{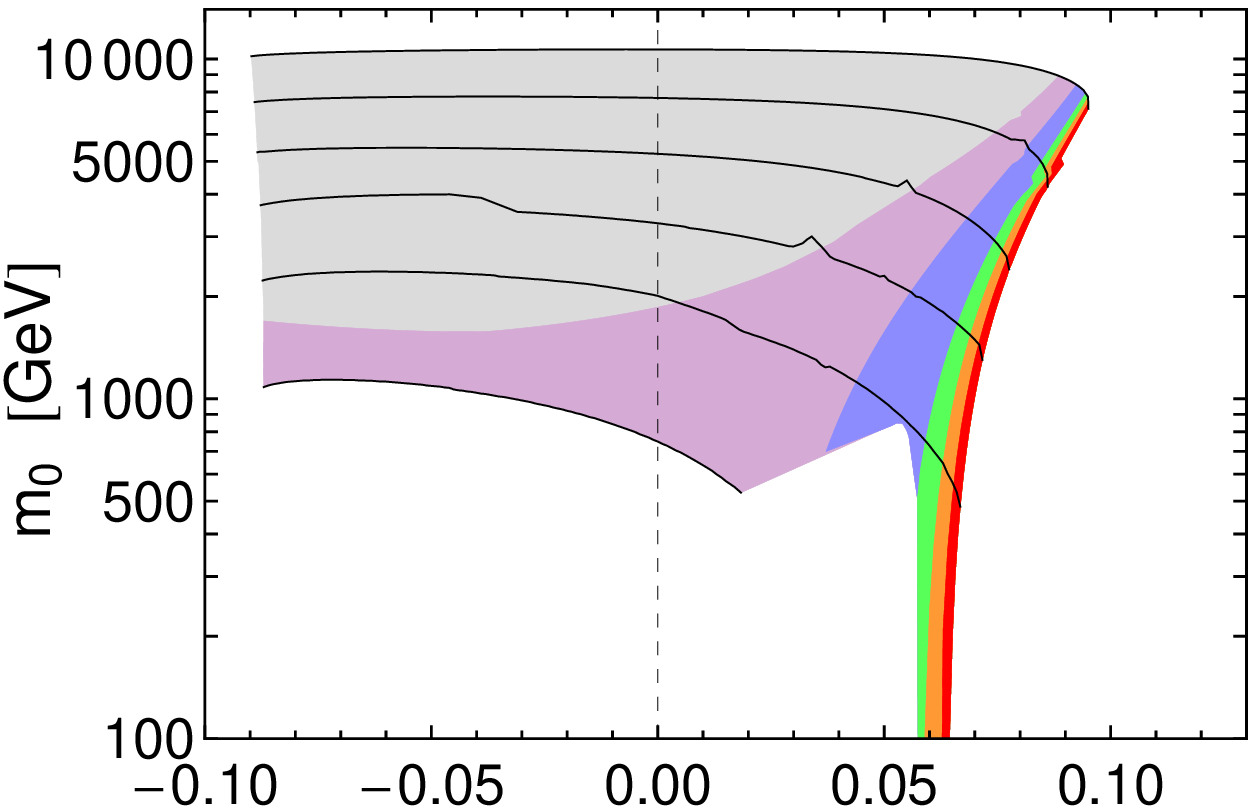}
 \end{minipage}
 \begin{minipage}[]{0.494\linewidth}
   \begin{flushright}
   \includegraphics[width=\linewidth,angle=0]{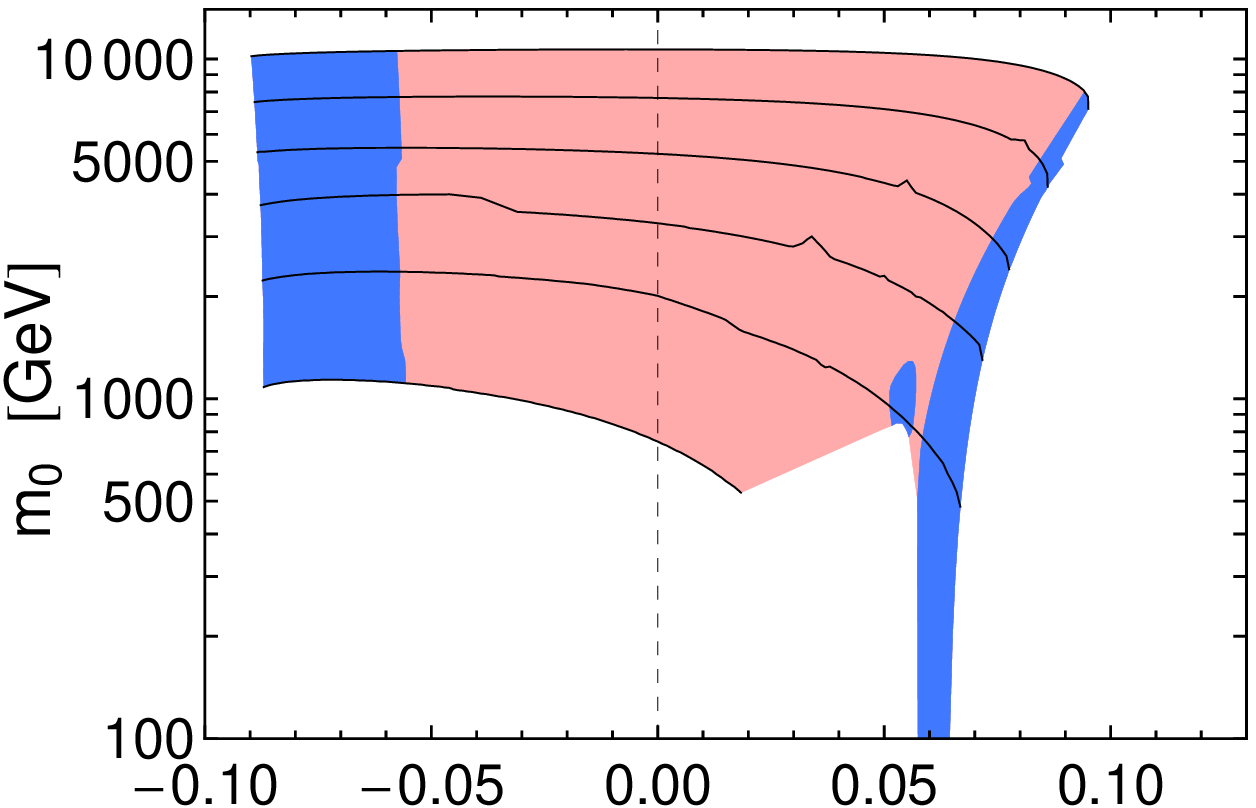}
   \end{flushright}
 \end{minipage}
 \begin{minipage}[]{0.494\linewidth}
   \includegraphics[width=\linewidth,angle=0]{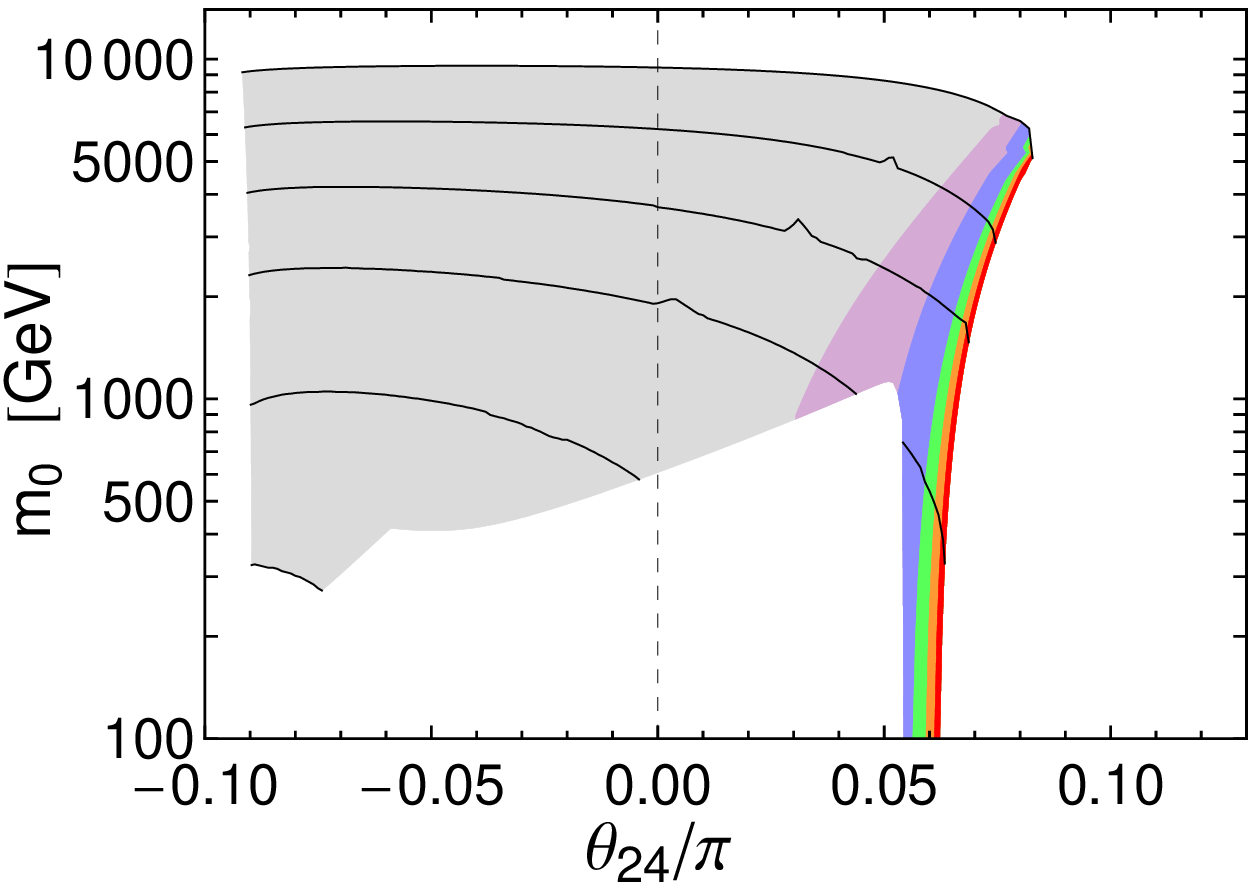}
 \end{minipage}
 \begin{minipage}[]{0.494\linewidth}
   \begin{flushright}
   \includegraphics[width=\linewidth,angle=0]{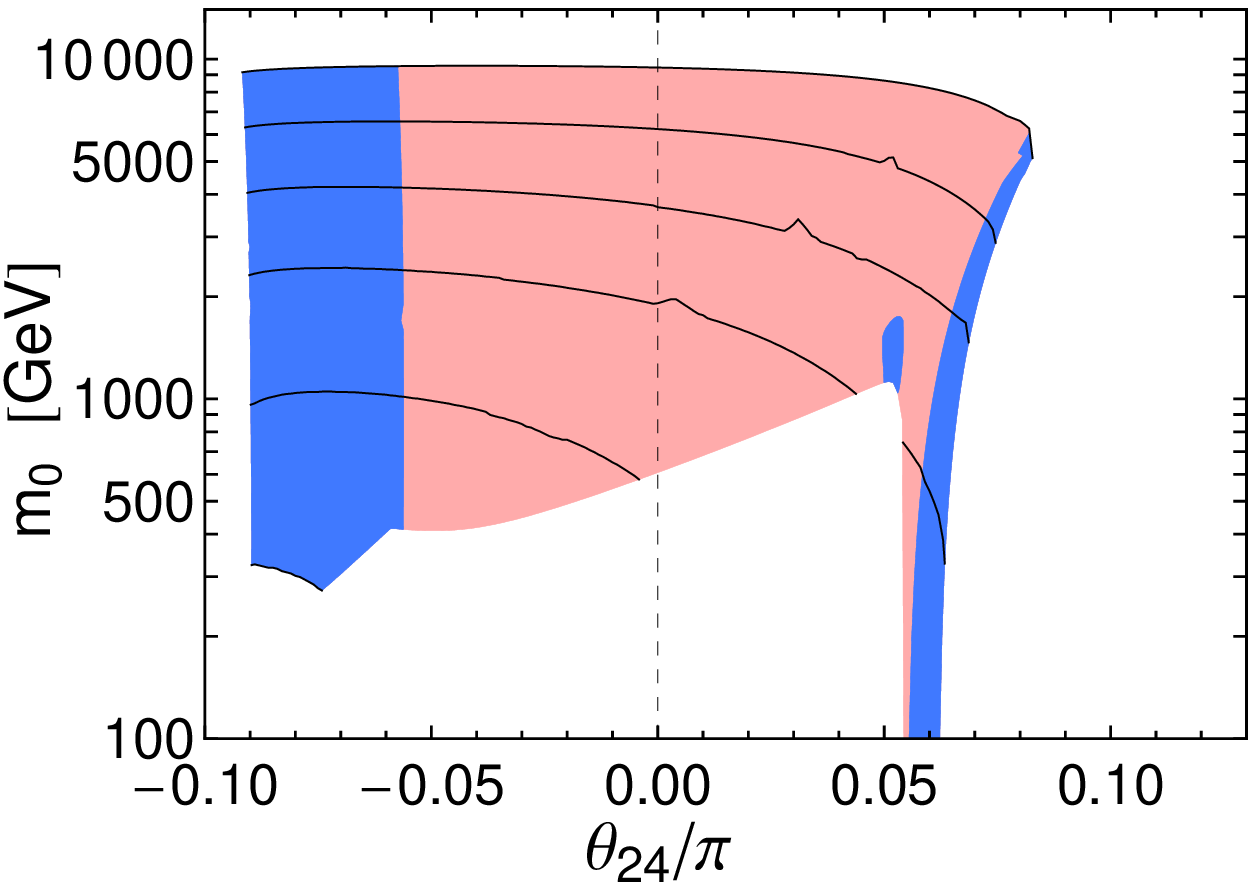}
   \end{flushright}
 \end{minipage}
\caption{\label{fig:tb20_Arat-1}
Maps of 
the $\mu$ parameter 
(left) and 
$\Omega_{\rm DM} h^2$ 
(right), as 
in Figure \ref{fig:tb10_Arat0_2000ALL}, but with 
$\tan\beta = 20$ and $A_0=-1$ and 
$M_3 = 600$, 1000, 1500, and 2000 GeV (from top to bottom) at $M_U$.} 
\end{figure}

Next, consider models in which the usual CMSSM-like parameter $A_0$ is 
constrained to be equal to $-m_0$ at $M_U$. This provides 
for a stronger mixing of top squarks, which in turn increases the 
prediction for $M_h$, allowing for models to be viable with lower overall 
superpartner masses. The maps of $\mu$ and $\Omega_{\rm DM} h^2$ in the $m_0$ vs. 
$\theta_{24}/\pi$ plane are shown for four choices $M_3 = 600$, $1000$, 
$1500$, and $2000$ GeV, in Figure \ref{fig:tb20_Arat-1}. Here I have fixed 
$\tan\beta = 20$.
The color scheme is the same as in the previous figures.

Note that for $A_0 = -m_0$, this time there is no focus point region with 
large $m_0$ and small $\mu$ that is consistent with the observed Higgs 
mass. (The focus point can be restored by generalizing it to include large 
non-universal scalar masses \cite{focuspoint6}.)
On the left side of the continent, there is again a region with 
allowed neutralino dark matter due to significant wino mixing in the LSP. 
However, this case is always associated with large $|\mu|$ in these 
models, and so could be viewed as disfavored for semi-natural 
supersymmetry and more generally in any motivational scheme in which 
$|\mu|$ is taken to be a proxy for fine-tuning. In the $M_3 = 1500$ and 
$2000$ GeV cases, we again see islands (centered near $\theta_{24}/\pi = 0.051$ and
$m_0 = 1300$ GeV in the latter case) of allowed neutralino dark matter 
associated with $\tilde N_1 \tilde C_1^\pm \rightarrow tb$, and $\tilde 
N_1 \tilde N_{2,3} \rightarrow b\overline b$ co-annihilations mediated by 
heavy Higgs bosons bringing about a reduction in the thermal relic 
abundance. These islands are significantly larger than in the $A_0=0$ case 
of the previous section, but are again associated with $\mu \gsim 1500$ 
GeV, with a mostly bino-like LSP that is significantly mixed with higgsinos that
are up to 250 GeV heavier.

Although there is no region with small $\mu$ or allowed neutralino dark 
matter in the CMSSM cases here (the vertical dashed lines in Figure 
\ref{fig:tb20_Arat-1}), the region with small $\mu$ along the right side 
of the continent (with $\theta_{24}/\pi \gsim 0.05$) persists, and in fact 
models with moderate and large $m_0$ are viable for much smaller gaugino 
masses than in the case of the previous section. This region extends down 
to very small $m_0$ when $M_3$ is larger than 1000 GeV at the GUT scale, 
again corresponding to a gaugino mass dominated scenario in which flavor 
violation is naturally suppressed. The region can provide allowed 
neutralino dark matter due to the significant higgsino content of the LSP, 
depending on how small $\mu$ is. Note that in the case of $M_3 = 2000$ GeV 
at $M_U$, the SuSpect prediction when $m_0$ is small is for $M_h$ between 
124 and 125 GeV, which may, however, actually be too high in view of the 
three-loop radiative corrections 
\cite{Martin:2007pg,Harlander:2008ju,Kant:2010tf,Feng:2013tvd}. For the 
$M_3 = 1500$ GeV figure, the SuSpect prediction is between 123 and 124 
GeV, and would therefore rely on 3-loop or other radiative corrections not 
included in SuSpect in order to bring it to the observed range. Values of 
$M_3$ between these two would interpolate between the two cases. As in 
Figure \ref{fig:tb10_Arat0}, in $M_3=1500$ and 2000 GeV in Figure 
\ref{fig:tb20_Arat-1} the lower bound of the allowed shaded region, when 
it is not a solid black line, is where the stau would become the LSP.


\section{Models with larger stop mixing ($A_0 = -2m_0$)\label{sec:Aminustwo}}
\setcounter{equation}{0}
\setcounter{figure}{0}
\setcounter{table}{0}
\setcounter{footnote}{1}

\begin{figure}[!tp]
 \begin{minipage}[]{0.494\linewidth}
   \includegraphics[width=\linewidth,angle=0]{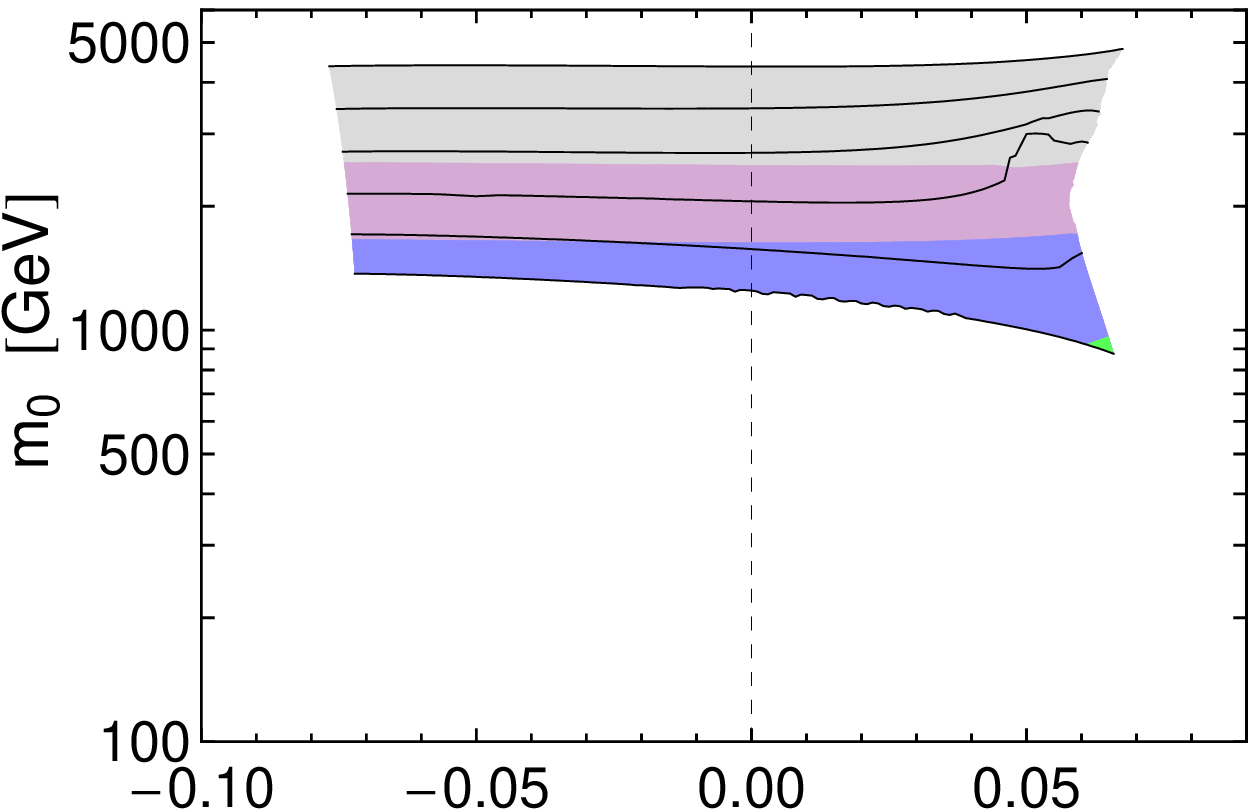}
 \end{minipage}
 \begin{minipage}[]{0.494\linewidth}
   \begin{flushright}
   \includegraphics[width=\linewidth,angle=0]{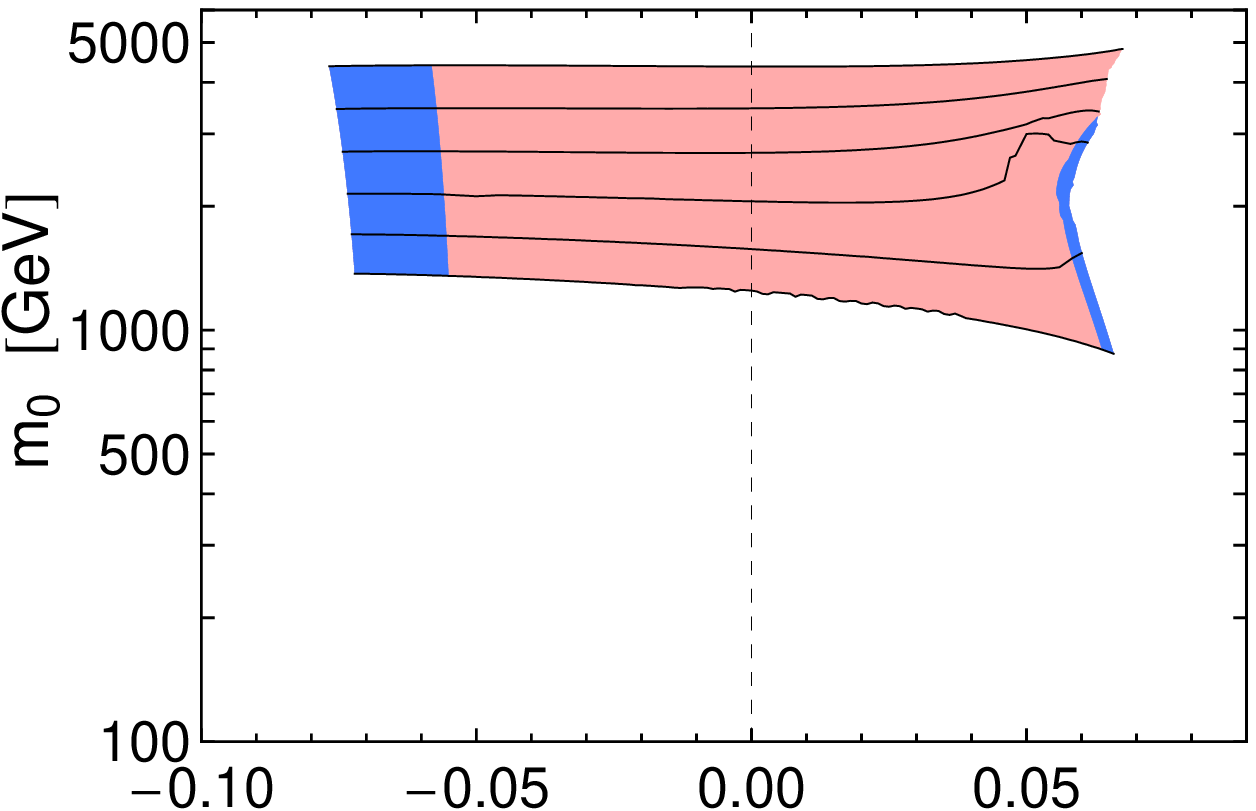}
   \end{flushright}
 \end{minipage}
 \begin{minipage}[]{0.494\linewidth}
   \includegraphics[width=\linewidth,angle=0]{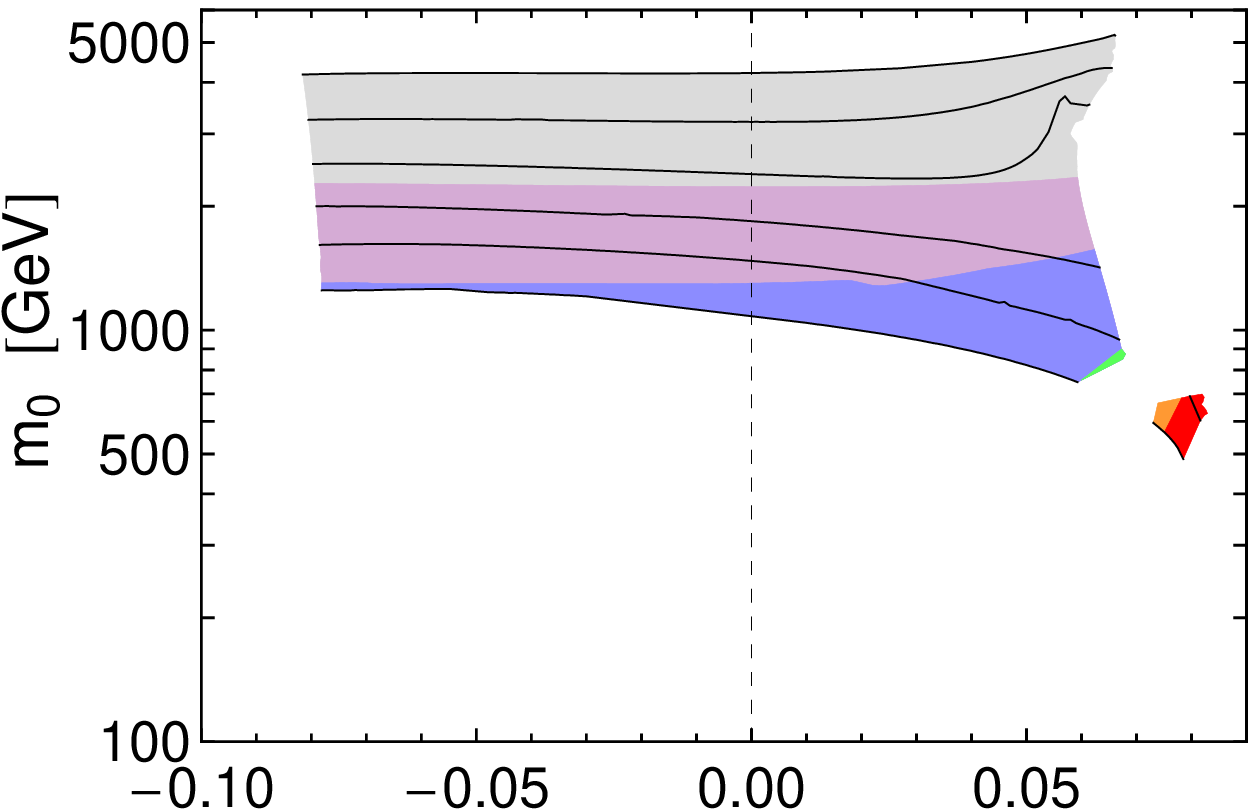}
 \end{minipage}
 \begin{minipage}[]{0.494\linewidth}
   \begin{flushright}
   \includegraphics[width=\linewidth,angle=0]{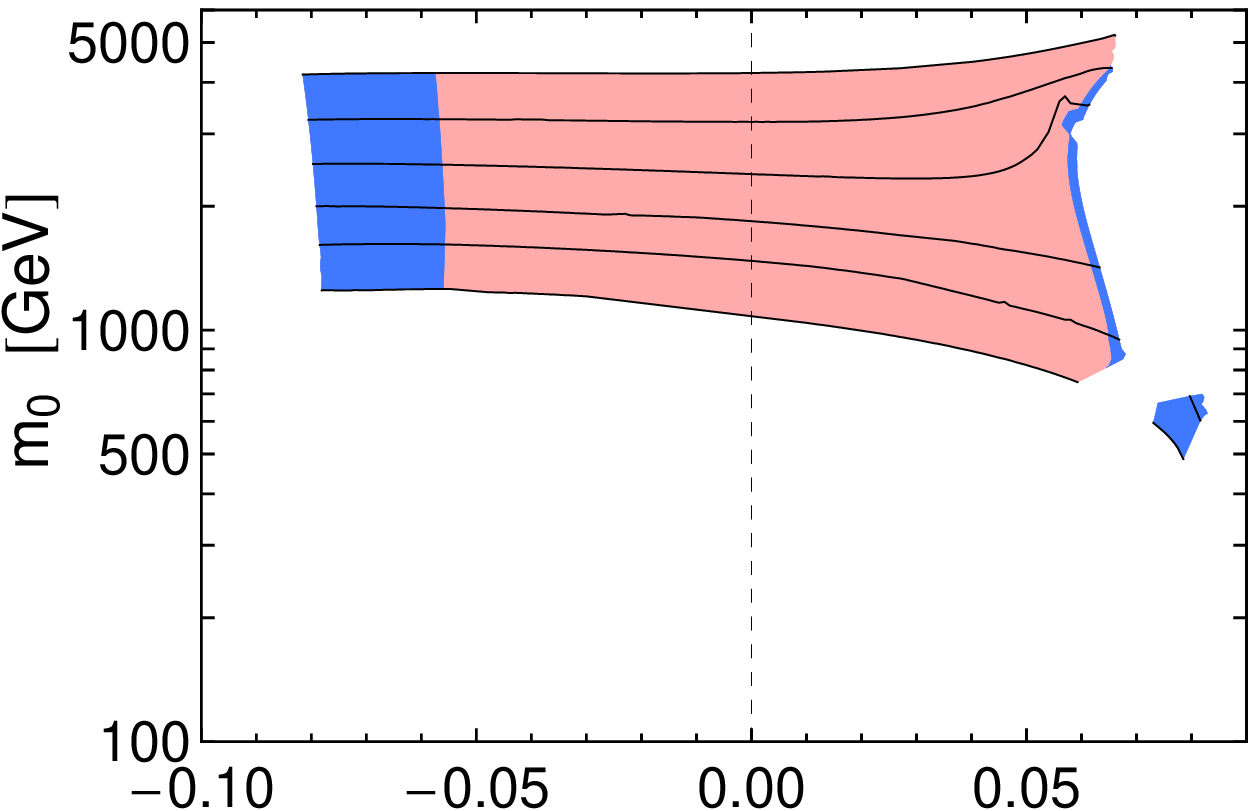}
   \end{flushright}
 \end{minipage}
 \begin{minipage}[]{0.494\linewidth}
   \includegraphics[width=\linewidth,angle=0]{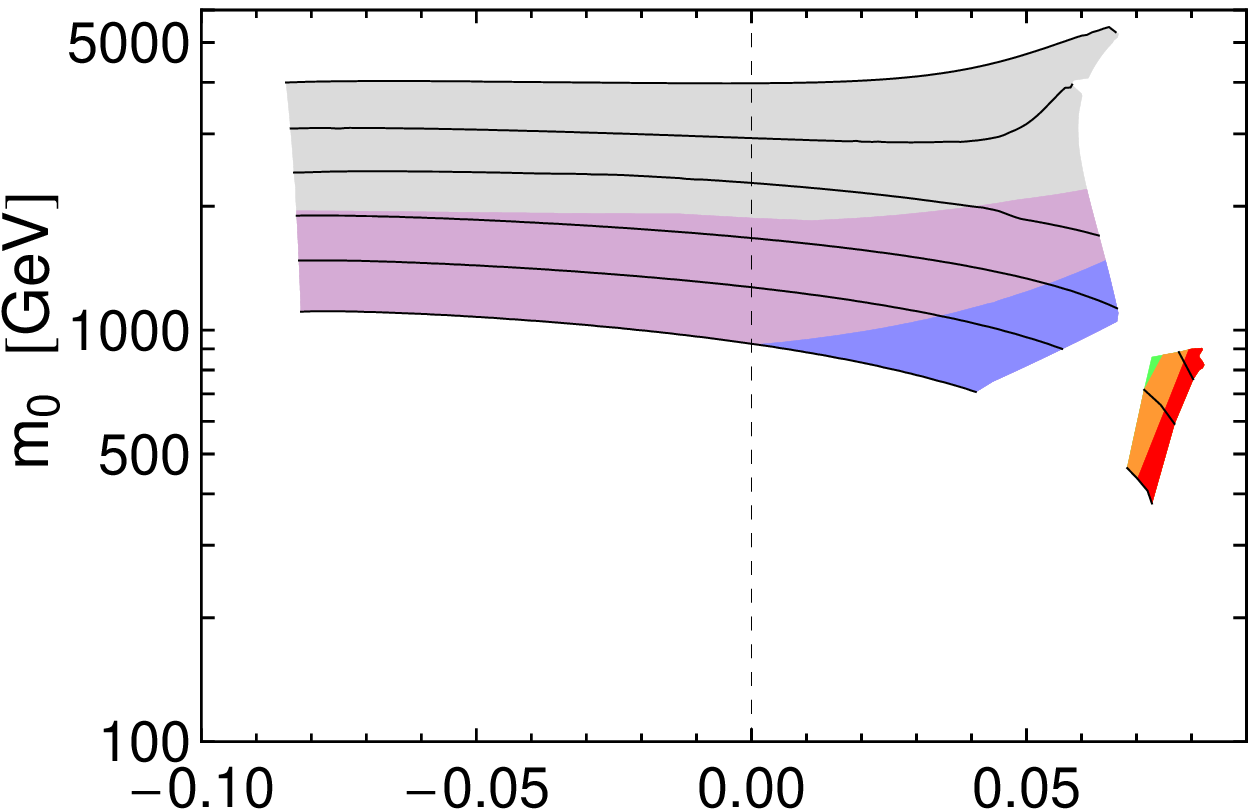}
 \end{minipage}
 \begin{minipage}[]{0.494\linewidth}
   \begin{flushright}
   \includegraphics[width=\linewidth,angle=0]{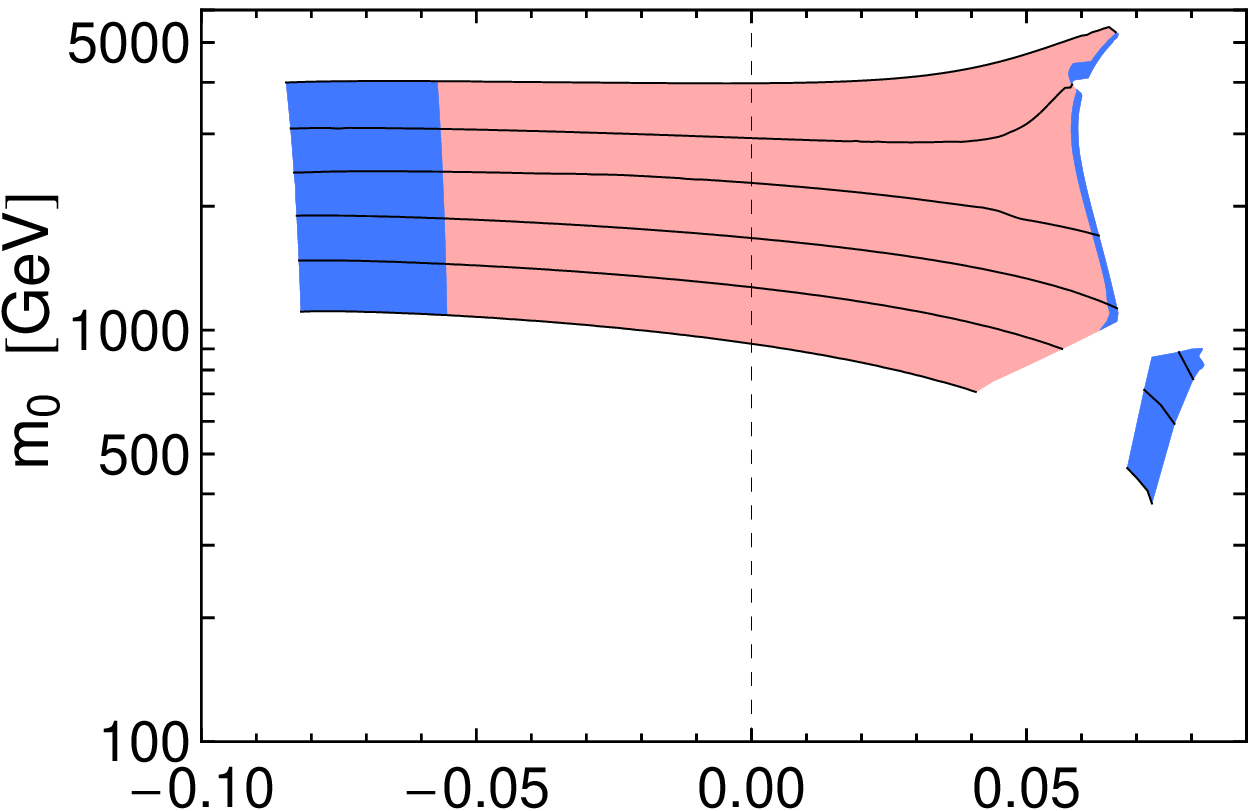}
   \end{flushright}
 \end{minipage}
 \begin{minipage}[]{0.494\linewidth}
   \includegraphics[width=\linewidth,angle=0]{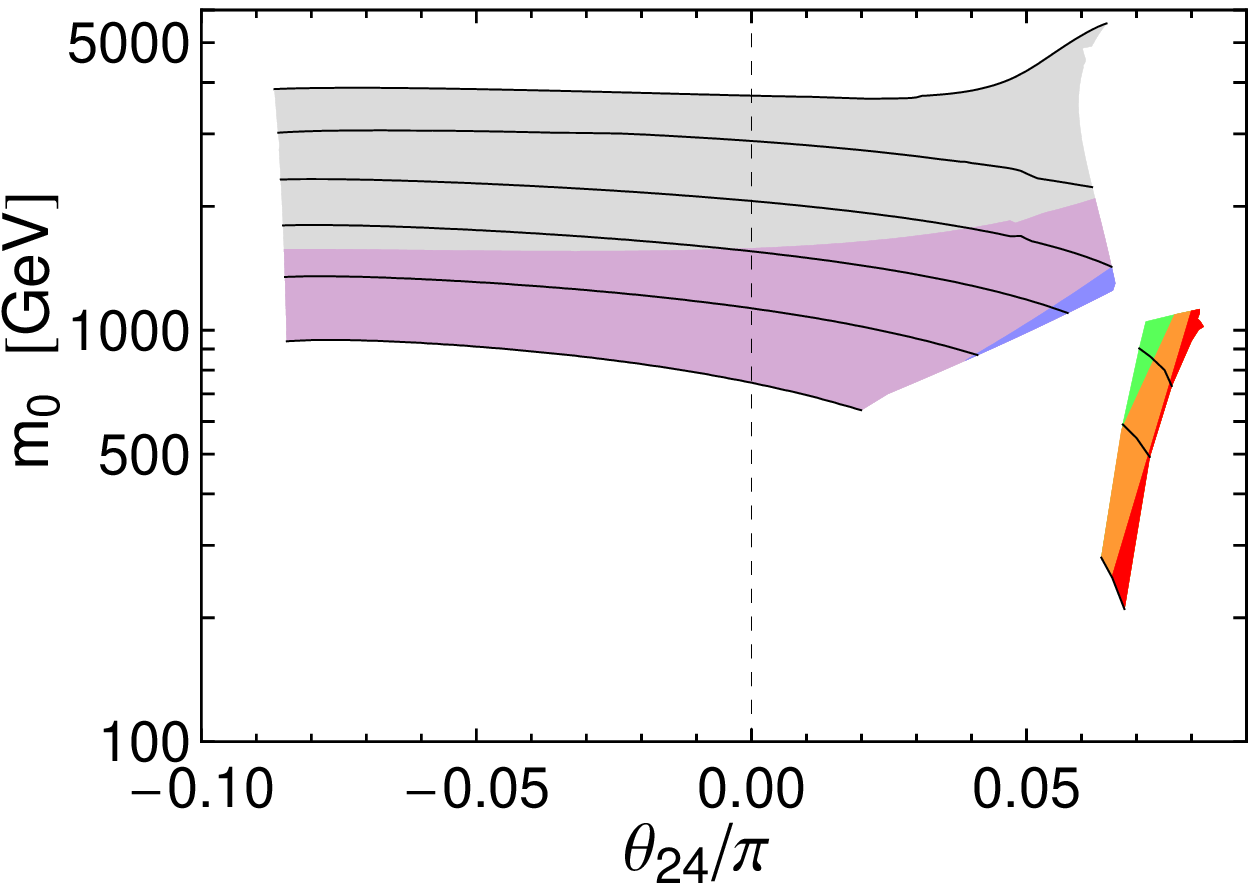}
 \end{minipage}
 \begin{minipage}[]{0.494\linewidth}
   \begin{flushright}
   \includegraphics[width=\linewidth,angle=0]{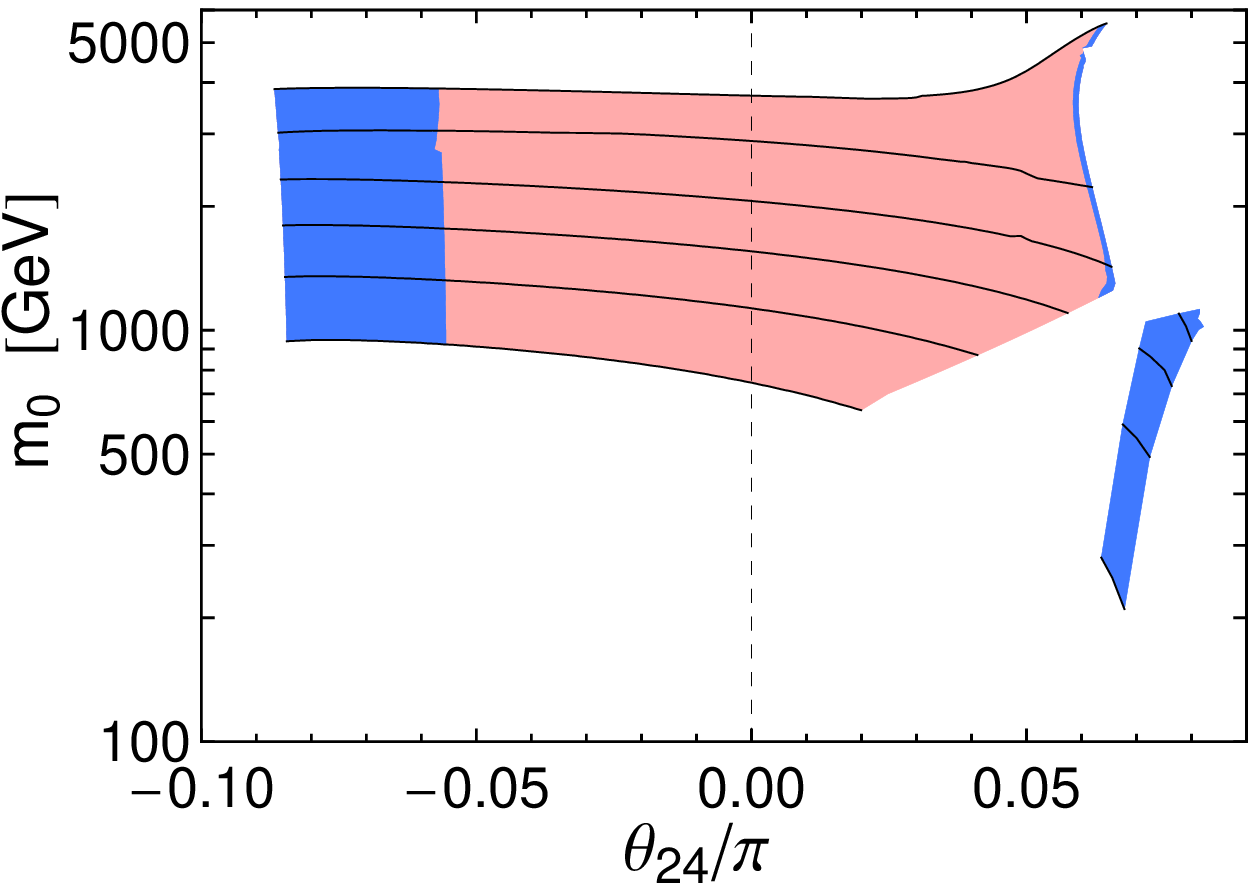}
   \end{flushright}
 \end{minipage}
\caption{\label{fig:tb30_Arat-2} 
Maps of 
the $\mu$ parameter 
(left) and 
$\Omega_{\rm DM} h^2$ 
(right), as 
in Figure \ref{fig:tb10_Arat0_2000ALL}, but with $\tan\beta = 30$ and $A_0=-2 m_0$ and 
$M_3 = 600$, 800, 1000, and 1200 GeV (from top to bottom) at $M_U$.} 
\end{figure}

In this section, I consider models with larger stop mixing, by taking 
$A_0 = -2m_0$ at $M_U$, so that $M_h$ is made larger. In 
figure \ref{fig:tb30_Arat-2}, I show the results of a map of this 
parameter space generalized to non-zero $\theta_{24}$, for $\tan\beta=10$, 
varying $m_0$ and four choices of $M_3 = 600$, 900, 1200, and 1500 GeV. 
Note that here we can accommodate the observed Higgs mass with much lower 
values of the gluino mass. The lowest value of 600 GeV was chosen because 
this gives a physical gluino mass of about $M_{\tilde g} = 1400$-1500 GeV 
(depending on the other parameters), and lower values would likely be 
ruled out by direct searches at the LHC, although such searches 
specifically for this scenario have not been conducted. Recent searches 
for supersymmetry in the CMSSM case at ATLAS have used as a comparison 
model the case with $\tan\beta = 30$ and 
$A_0/m_0 = -2$ rather than 0. This allows the CMSSM models to accommodate 
$M_h=126$ GeV, unlike the old standard choices of $\tan\beta = 10$ and 
$A_0=0$.  A similar set of maps is therefore shown in Figure 
\ref{fig:tb30_Arat-2} again for $A_0/m_0 = -2$ but with $\tan\beta=30$. For 
the CMSSM $\theta_{24} = 0$ case, the ATLAS limit 
\cite{ATLASSUSY1,ATLASSUSY2,ATLASSUSY3} is 
between $M_3 = 550$ and 800 GeV, depending on $m_0$.
\begin{figure}[!tp]
 \begin{minipage}[]{0.494\linewidth}
   \includegraphics[width=\linewidth,angle=0]{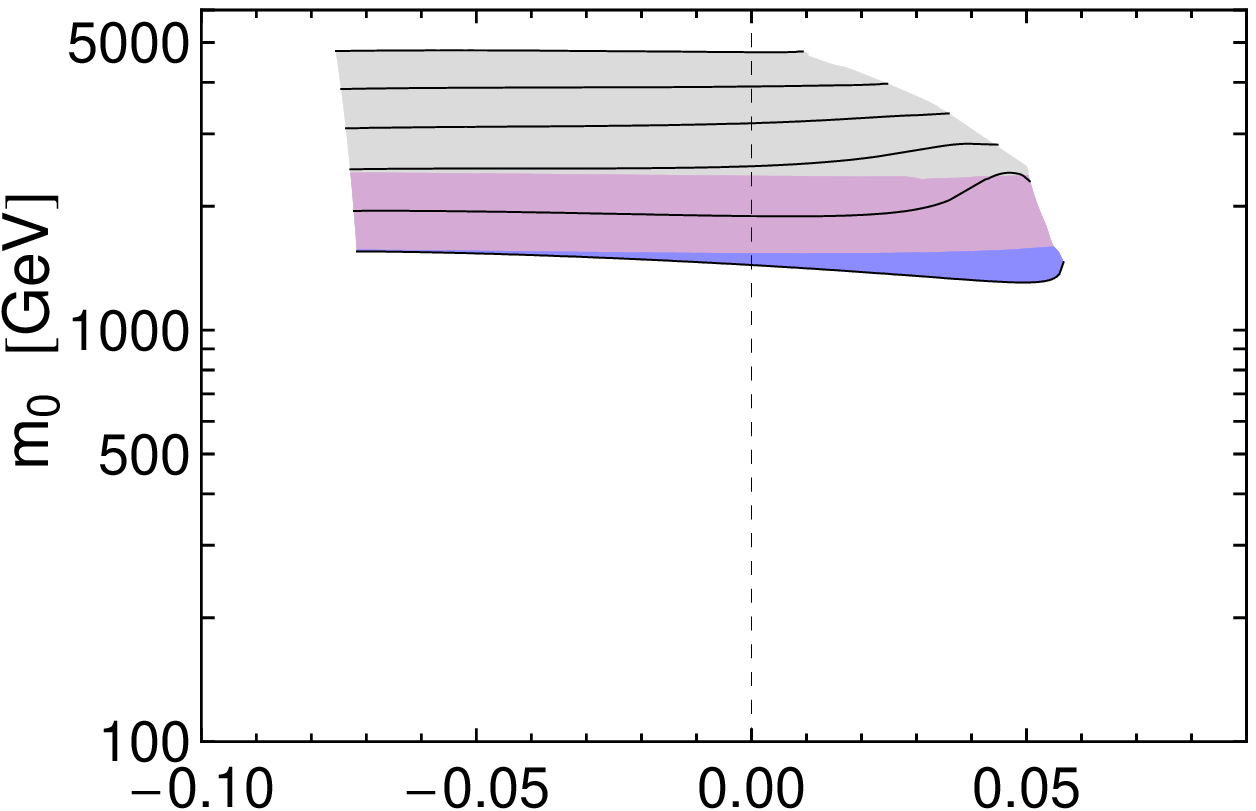}
 \end{minipage}
 \begin{minipage}[]{0.494\linewidth}
   \begin{flushright}
   \includegraphics[width=\linewidth,angle=0]{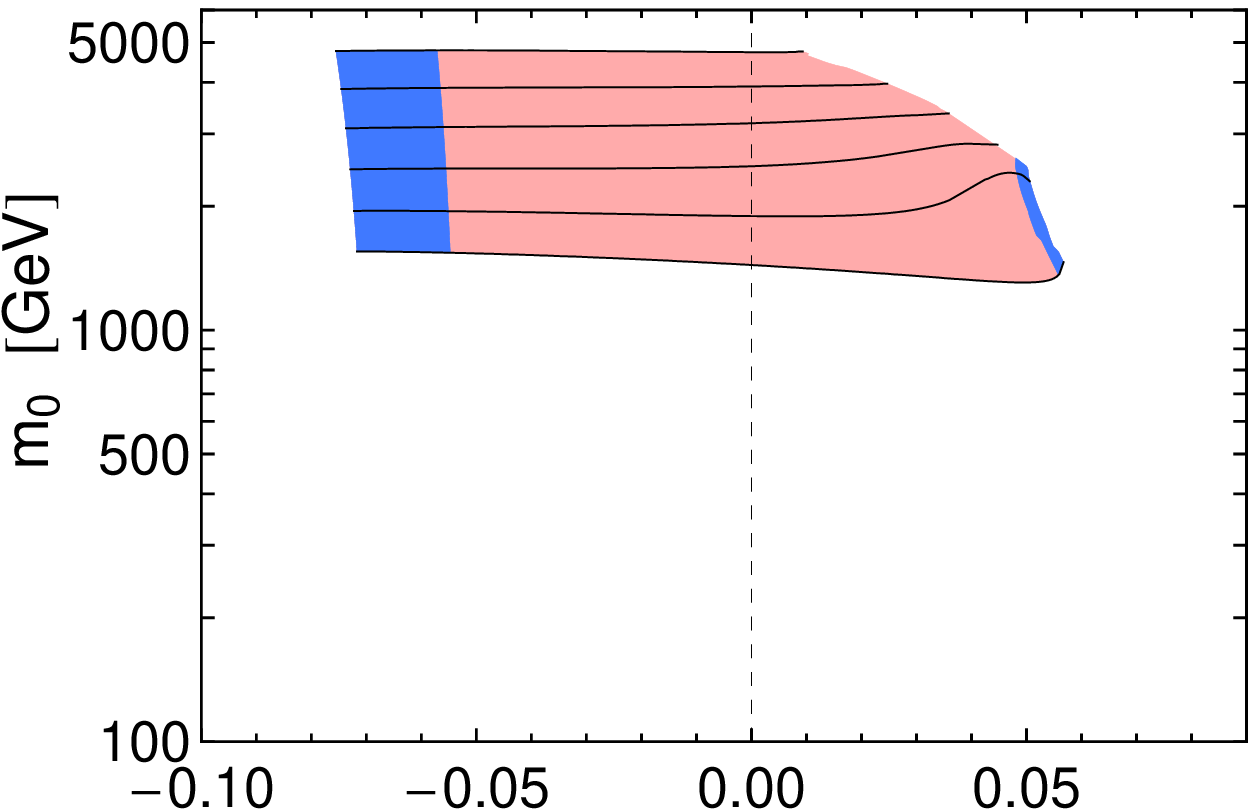}
   \end{flushright}
 \end{minipage}
 \begin{minipage}[]{0.494\linewidth}
   \includegraphics[width=\linewidth,angle=0]{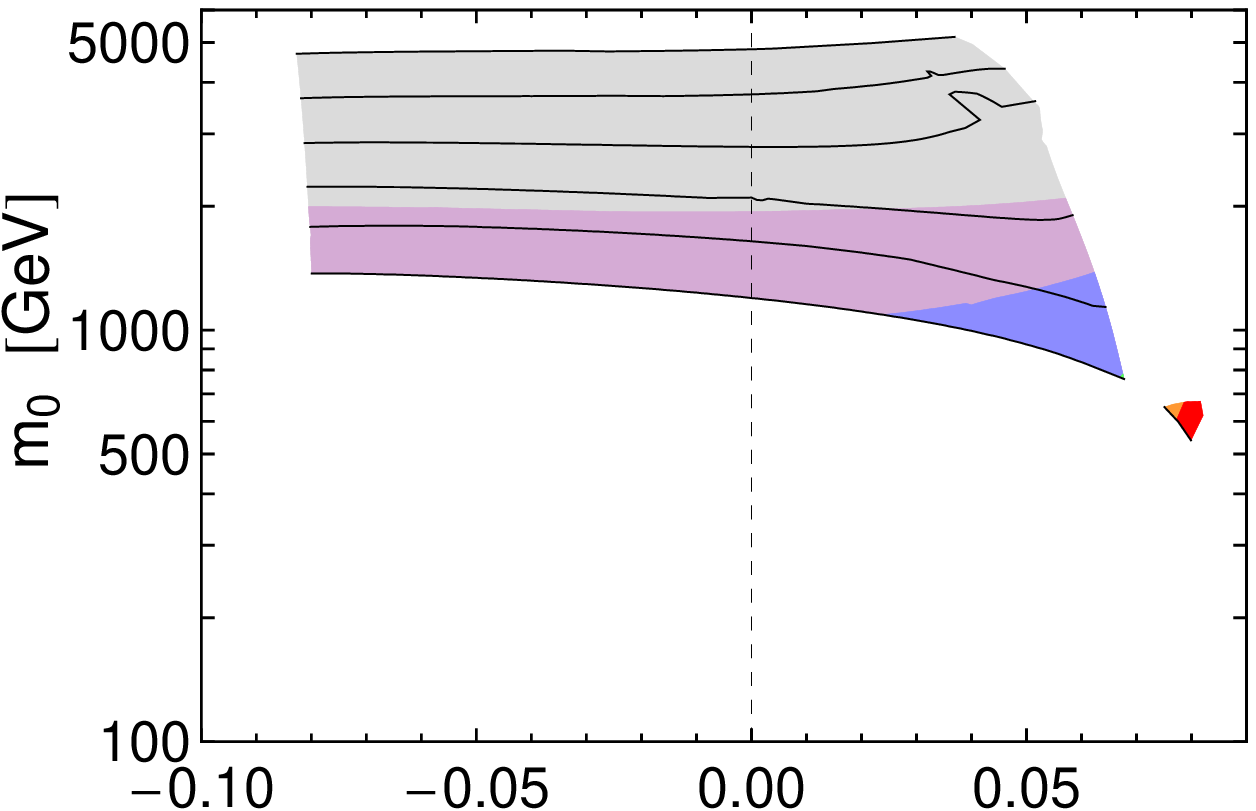}
 \end{minipage}
 \begin{minipage}[]{0.494\linewidth}
   \begin{flushright}
   \includegraphics[width=\linewidth,angle=0]{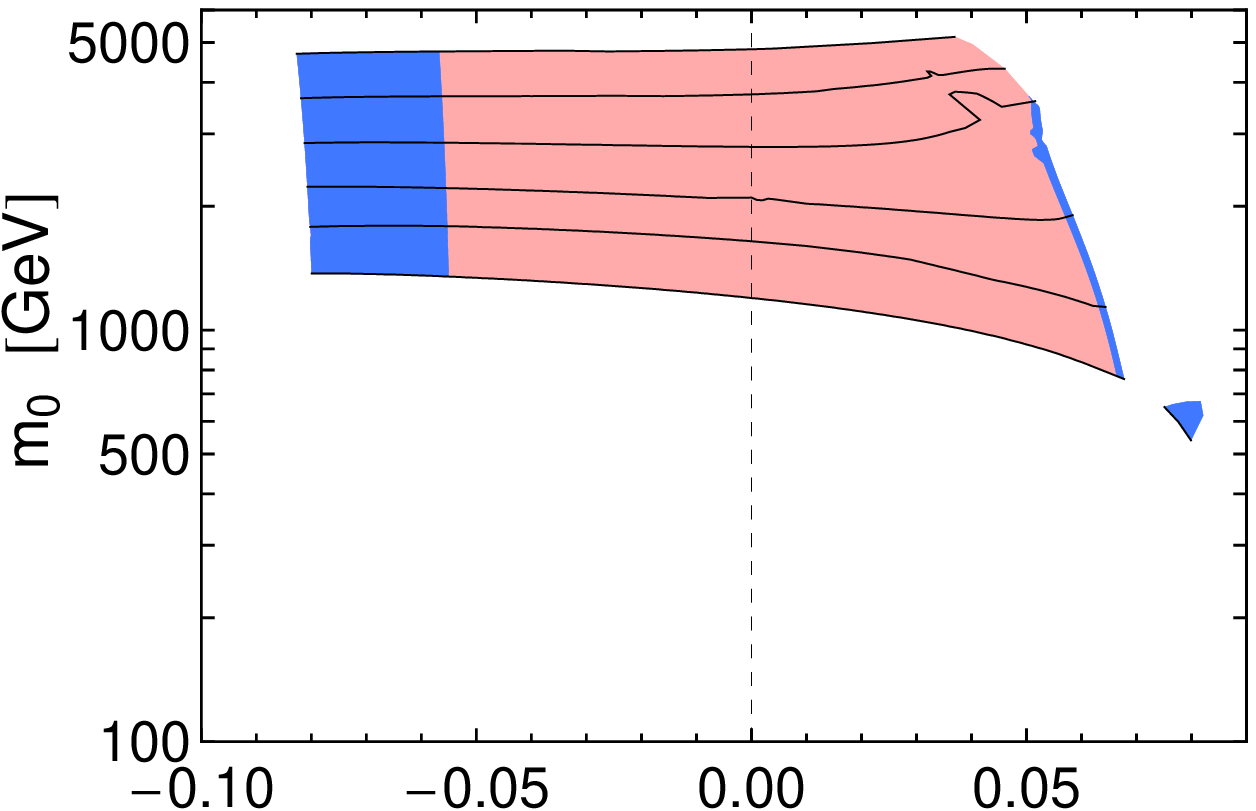}
   \end{flushright}
 \end{minipage}
 \begin{minipage}[]{0.494\linewidth}
   \includegraphics[width=\linewidth,angle=0]{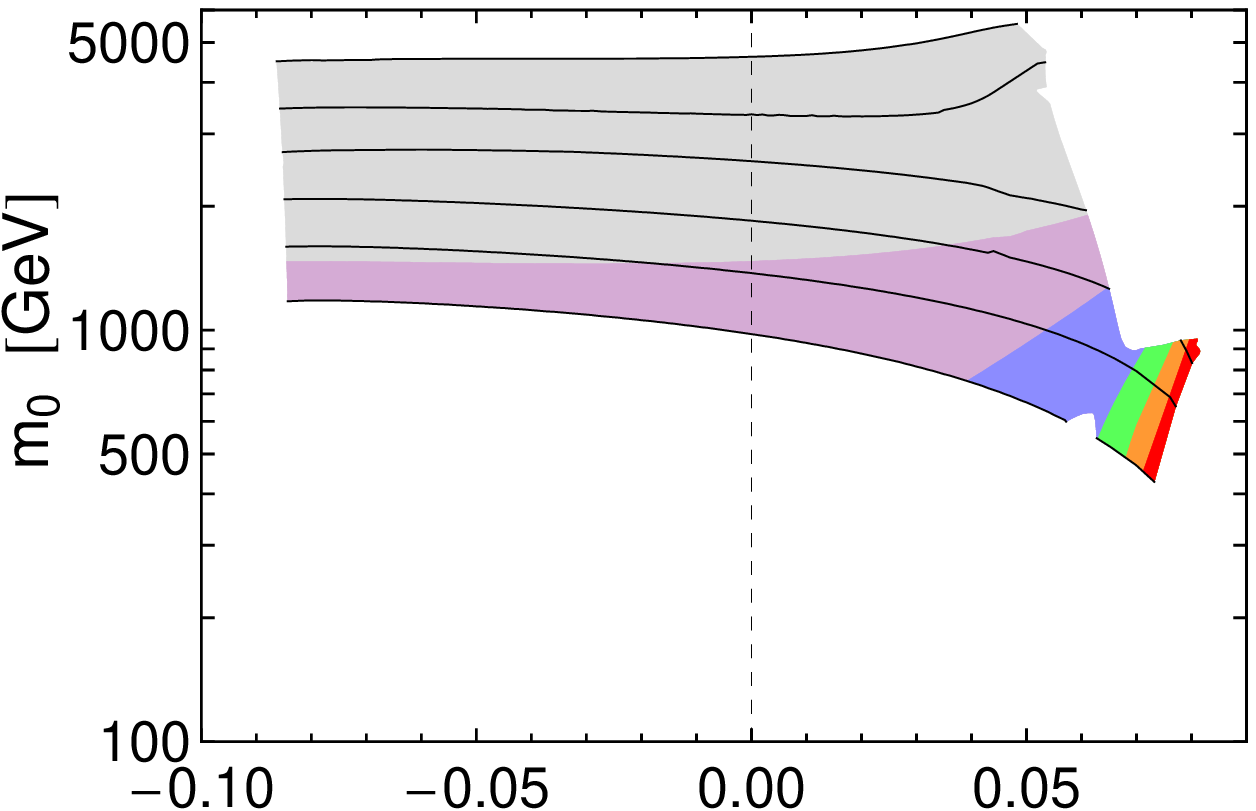}
 \end{minipage}
 \begin{minipage}[]{0.494\linewidth}
   \begin{flushright}
   \includegraphics[width=\linewidth,angle=0]{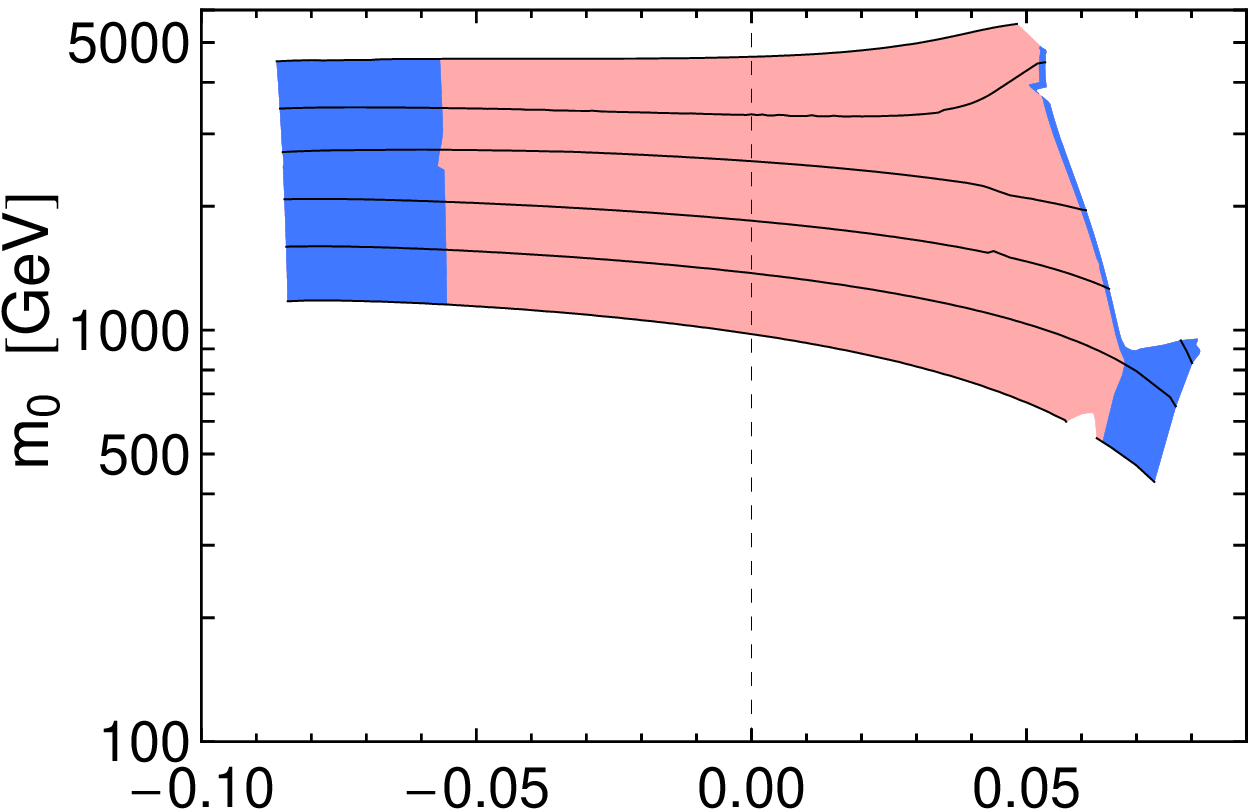}
   \end{flushright}
 \end{minipage}
 \begin{minipage}[]{0.494\linewidth}
   \includegraphics[width=\linewidth,angle=0]{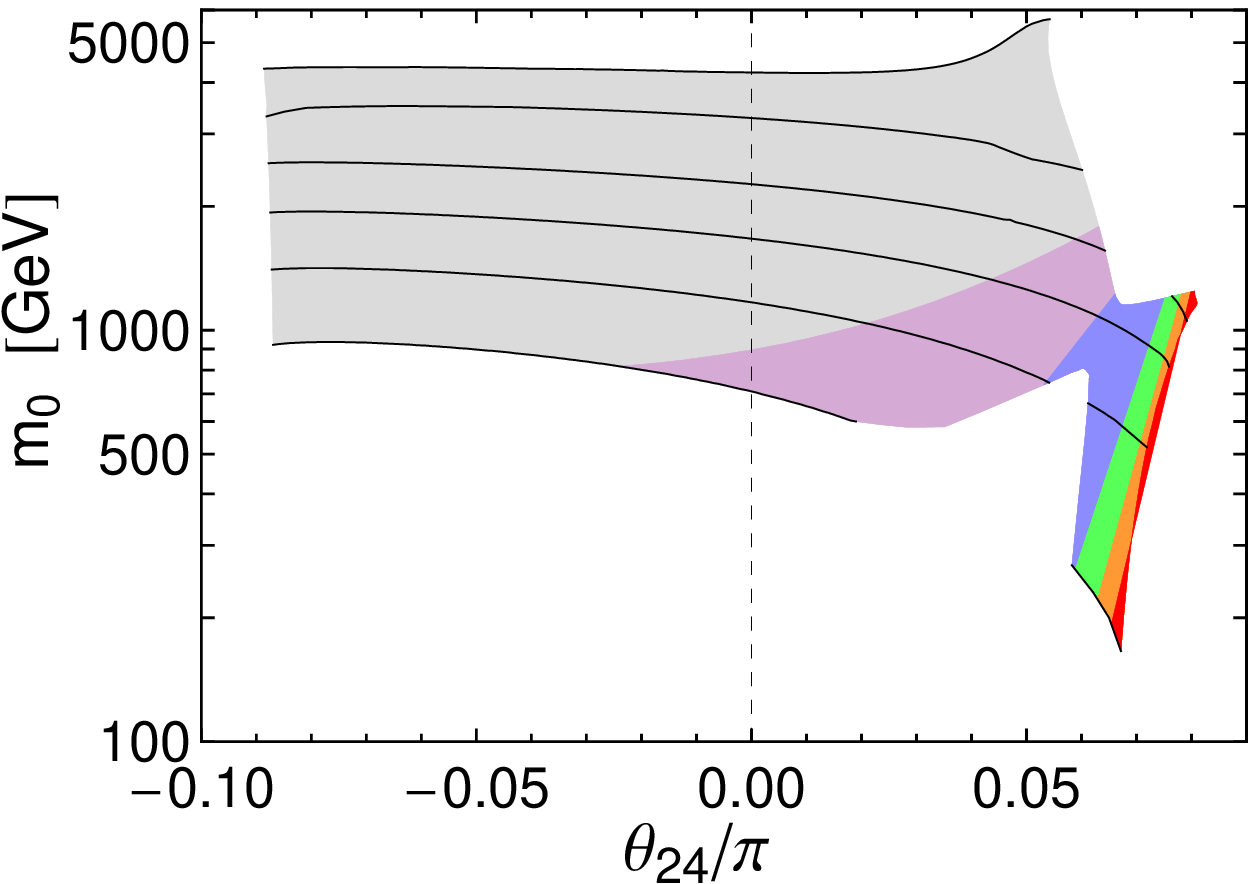}
 \end{minipage}
 \begin{minipage}[]{0.494\linewidth}
   \begin{flushright}
   \includegraphics[width=\linewidth,angle=0]{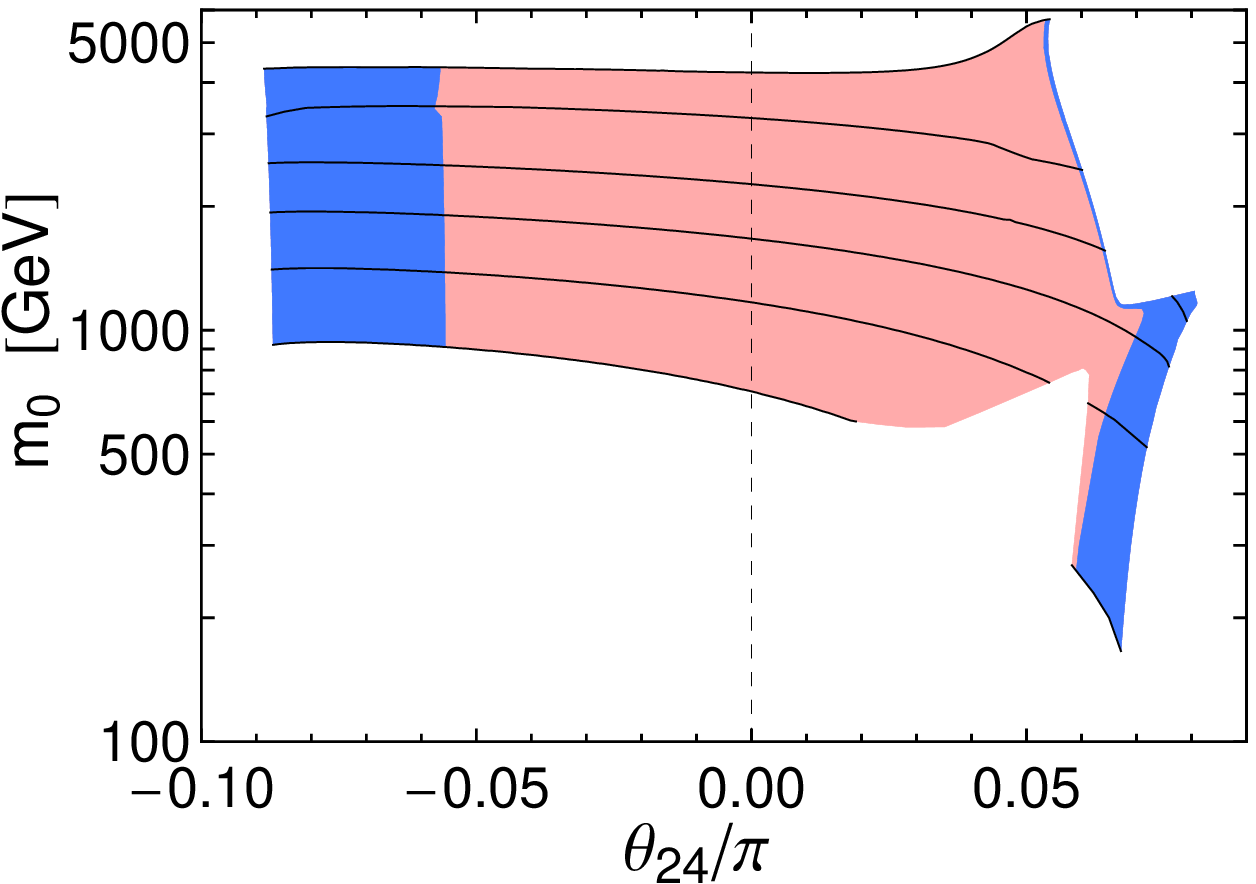}
   \end{flushright}
 \end{minipage}
\caption{\label{fig:tb10_Arat-2}
Maps of 
the $\mu$ parameter 
(left) and 
$\Omega_{\rm DM} h^2$ 
(right), as 
in Figure \ref{fig:tb10_Arat0_2000ALL}, but with $\tan\beta = 10$ and $A_0=-2 m_0$ and 
$M_3 = 600$, 900, 1200, and 1500 GeV (from top to bottom) at $M_U$.
} 
\end{figure}

The maps shown in Figures \ref{fig:tb10_Arat-2} and \ref{fig:tb30_Arat-2} 
only go up to $M_3 = 1500$ and 1200 GeV respectively, to emphasize that 
this is all that is needed in order to accommodate the observed Higgs boson 
mass. However, there is nothing wrong with 
higher $M_3$, which would allow even lower $m_0$, 
and increase the regions where semi-natural supersymmetry with small $\mu$ 
is allowed. In fact, for $M_3$ larger than about 1500 GeV, the results for small 
$m_0$ are very similar to those of the previous sections, because then 
$A_0$ will also be small compared to the overall scale set by gaugino masses,
due to the assumed relation $A_0 = -2 m_0$. 
Therefore, plots with larger $M_3$ are not shown, for 
brevity. Note that the maps of Figures \ref{fig:tb10_Arat-2} and 
\ref{fig:tb30_Arat-2} are also qualitatively similar to the $A_0=-m_0$ 
maps of the previous section, in many respects, even for the small $M_3$ 
values shown. In particular, there is no focus point region at large 
$m_0$, but there is a small $\mu$ region for $\theta_{24}/\pi = 0.055$ to 
$0.08$. The region is confined to $m_0 \lsim 1200$ GeV here, because for 
larger $A_0=-2m_0$ the lighter top squark decreases in mass due to the 
top-squark mixing, and becomes the LSP. The thinner strips of allowed dark 
matter along the upper right-hand sides of the shaded allowed regions in 
Figures \ref{fig:tb10_Arat-2} and \ref{fig:tb30_Arat-2} are stop 
co-annihilation \cite{stopco1,stopco2,stopco3,stopco4} 
and stop-mediated annihilation 
\cite{compressedSUSY,compressedSUSY2} regions, where the top squark is not 
much heavier than the LSP. 
The lower boundary wedges biting into the shaded allowed regions in Figure 
\ref{fig:tb10_Arat-2} for $M_3 = 1200$ and 1500 GeV are where the lighter 
stau is the LSP. In the $M_3 = 900$ GeV case of Figure 
\ref{fig:tb10_Arat-2}, the regions vetoed by
stop LSP and stau LSP collide, 
separating the small $\mu$ region off into an island. The same thing is 
responsible for the islands in Figure \ref{fig:tb30_Arat-2} for $M_3 = 
800$, 1000, and 1200 GeV. Throughout these islands, one finds $\Omega_{\rm DM} h^2 
\lsim 0.12$ due to a large higgsino content of the LSP.
 

\section{Superpartner mass spectra and discovery 
prospects\label{sec:directdetection}}
\setcounter{equation}{0}
\setcounter{figure}{0}
\setcounter{table}{0}
\setcounter{footnote}{1}

In this section, I will discuss the specifics of the superpartner mass 
spectrum for certain models discussed above. There are two distinct 
well-motivated branches of parameter space that I will consider. First, 
one can take semi-naturalness as the main motivation, and require 
$|\mu|$ less than a few hundred GeV. In this case, the predicted thermal 
$ \Omega_{\rm DM} h^2$ is much less than $0.12$, so then I will simply 
assume that axions (or some other particles) are the dark matter, and 
apply no constraints from direct detection. Second, one can require 
instead that $\Omega_{\rm DM} h^2 = 0.12$, so that the neutralino LSP is 
the dark matter, with the correct thermal relic abundance. This requires 
larger values of $\mu$, but typically still in the range of about 1000 
GeV. Here one can apply constraints from dark matter direct detection 
experiments.

\begin{figure}[!tp]
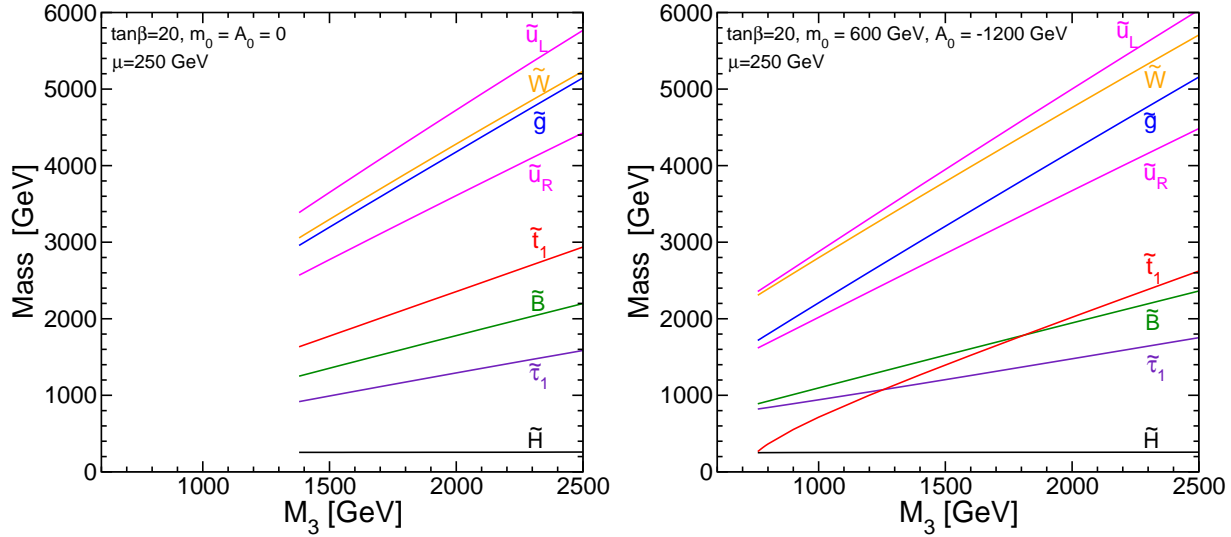

 \begin{minipage}[]{0.49\linewidth}
   \includegraphics[width=0.99\linewidth,angle=0]{spectrum_0_0_20.eps}
 \end{minipage}
 \begin{minipage}[]{0.49\linewidth}
   \includegraphics[width=0.99\linewidth,angle=0]{spectrum_-2_600_20.eps}
 \end{minipage}
\caption{\label{fig:spectrum}
Mass spectra for selected superpartners, as a function of $M_3$ at 
$M_U$, for $\tan\beta = 20$, with $m_0 = A_0 = 0$ (left panel) 
and $m_0 = -A_0/2 = 600$ GeV (right panel). In each case, the gaugino mass 
non-universality parameter $\theta_{24}$ is determined by requiring $\mu = 
250$ GeV. The lines labeled by $\tilde H$ indicate the nearly degenerate 
$\tilde N_1$, $\tilde C_1$, and $\tilde N_2$, while the lines labeled by 
$\tilde W$ indicate the nearly degenerate $\tilde C_2, \tilde N_4$, and 
$\tilde B$ is $\tilde N_3$. Below the lower endpoints of the lines in the 
left panel, SuSpect predicts $M_h < 123$ GeV. In the right panel, the 
lower endpoints are set by requiring that the LSP not be a top squark.}
\end{figure}

In the first, ``semi-natural", supersymmetry case, which is defined for 
practical purposes to include the region with $\theta_{24} \gsim 0.05$ and 
small $m_0^2$ at $M_U$, a typical one-parameter 
superpartner mass spectrum is shown in Figure 
\ref{fig:spectrum}a. Here I have taken $\tan\beta = 20$ and $m_0 = A_0 = 
0$ GeV as in ``no-scale" CMSSM models, and varied $M_3$ at $M_U$, 
with the gaugino mass non-universality parameter
$\theta_{24}$ fixed for each model point so that $\mu=250$ GeV.
The left cutoff of the lines is fixed by demanding $M_h > 123$ GeV 
according to SuSpect. The lightest superpartners are two mostly 
higgsino-like neutralinos and a chargino, with masses given at tree level 
by the approximate formulas:
\beq
M_{\tilde N_1} &=& 
|\mu| - \frac{m_W^2 [\pm 1 + \sin(2\beta)] 
[M_1 + M_2 \tan^2\theta_W\, - \mu/\cos^2\theta_W]}{2 (M_1 - \mu)(M_2 - \mu)} 
+ \ldots
\\
M_{\tilde C_1} &=&
|\mu| \mp \frac{m_W^2 [\mu  + M_2 \sin(2 \beta)]}{M_2^2 - \mu^2} + \ldots 
\\
M_{\tilde N_2} &=& 
|\mu| + \frac{m_W^2 [\pm 1 - \sin(2\beta)] 
[M_1 + M_2 \tan^2\theta_W\, + \mu/\cos^2\theta_W]}{2 (M_1 + \mu)(M_2 + \mu)} 
+ \ldots
\eeq
where electroweak symmetry breaking is treated as a perturbation, and $\pm 
1$ is the sign of $\mu$. The higgsino-like states satisfy $M_{\tilde N_1} < 
M_{\tilde C_1} < M_{\tilde N_2}$ for positive $\mu$, with a mass splitting that decreases as 
$M_1$ and/or $M_2$ are taken large compared to $m_W$ and $|\mu|$. 
(In the specific model 
line shown, the total mass splitting of the higgsino-like states,
$m_{\tilde N_2}- m_{\tilde N_1}$, is a few GeV. One-loop radiative corrections
to the tree-level formulas are significant, but can mostly be absorbed into
the definition of the scale-dependent parameter $\mu$.) 
There is nothing special about the value of $\mu=250$ GeV chosen here; it 
could even be as low as the current bounds from LEP2 of close to 100 GeV. 
The LHC discovery potential for this whole class of models is 
the same as discussed already in \cite{Baer:2011ec}, and is
extremely challenging. (See also \cite{Han:2013usa}. 
A similar case with nearly degenerate
gauginos rather than higgsinos is discussed in \cite{Gori:2013ala}.) 
The direct production 
cross-section for higgsinos is quite small, and the decay products will 
give rise to soft leptons and jets, with significant physics and detector 
backgrounds and low trigger efficiencies, so that discovery may have to rely
on extra radiated jets. 
Unlike the cases studied in \cite{Baer:2013yha,Han:2013kza}, 
there are no superpartners that are close in mass that could decay to the 
higgsinos; in particular, the very motivation of this class of 
semi-natural supersymmetry models ensures that the wino-like state masses will 
be around the gluino mass, and 
much too heavy to produce with an appreciable cross-section at the LHC. 
From Figure \ref{fig:spectrum} we see that the discovery of the gluino and 
squarks is also quite problematic for the LHC, unless their masses are 
near the lower end of the range shown. It may also be possible to search for
the light higgsino-like states using weak boson fusion 
\cite{WBF1,WBF2,WBF3}.

A more optimistic model line is shown in Figure \ref{fig:spectrum}b, where 
now I have chosen $m_0 = 600$ GeV and $A_0 = -1200$ GeV in order to 
accommodate $M_h$ consistently with lower superpartner masses. Again I have 
fixed $\tan\beta=20$ and adjusted $\theta_{24}$ for each model point so 
that $\mu = 250$ GeV, providing for light higgsino-like states. In this 
case, the lower endpoint of the model line is given by the point at which 
a top squark becomes the LSP, and in this case the gluino and right-handed 
squarks could perhaps already be light enough to detect with the current data at 
the LHC with $\sqrt{s} = 8$ TeV.

In both of the above cases, the heavier higgsino-like states will 
decay through off-shell weak bosons to the LSP according to
\beq
\tilde C_1 \>\rightarrow\> W^{(*)} \tilde N_1,
\qquad\qquad
\tilde N_2 \>\rightarrow\> Z^{(*)} \tilde N_1,
\eeq
with branching ratios that are distorted by kinematics to 
disfavor tau and charm final states, to an extent that depends
on the mass differences. The lighter top squark will decay according to
\beq
\tilde t_1 \rightarrow
\left \{ \begin{array}{lll}
b \tilde C_1\qquad(\sim 50\%),
\\[-4pt]
t \tilde N_2\qquad(\sim 25\%),
\\[-4pt]
t \tilde N_1\qquad (\sim 25\%),
\end{array}
\right.
\label{eq:stopdecays}
\eeq
for $m_{\tilde t_1} - m_{\tilde N_1} \gg m_t$, and
\beq
\tilde t_1 \rightarrow b \tilde C_1\qquad(100\%)
\eeq
for $m_{\tilde t_1} - m_{\tilde N_1} < m_t$, and branching ratios that are 
intermediate between these extremes if $m_{\tilde t_1} - m_{\tilde N_1}$ 
is comparable to but larger than $m_t$. The lighter up and down quarks 
decay according to 
$\tilde q_R \rightarrow q \tilde N_3$ with a nearly 100\% branching ratio, 
while the gluino has its largest decay branching fractions equally to $t 
\overline{\tilde t_1}$ and $\overline t \tilde t_1$. 
The bino-like neutralino $\tilde N_3$ also has large branching fractions to
$t\overline{\tilde t_1}$ and $\overline t \tilde t_1$, but also to 
$W^\pm \tilde C_1^\mp$.
Thus the usual 
supersymmetry signals apply for gluino and squark production, but only if 
they are light enough to be produced in sufficient numbers at the LHC.
Note that the wino-like states $\tilde C_2, \tilde N_4$ are heavier 
than the gluino and so
will not participate in LHC discovery signals.

\begin{figure}[!tp]
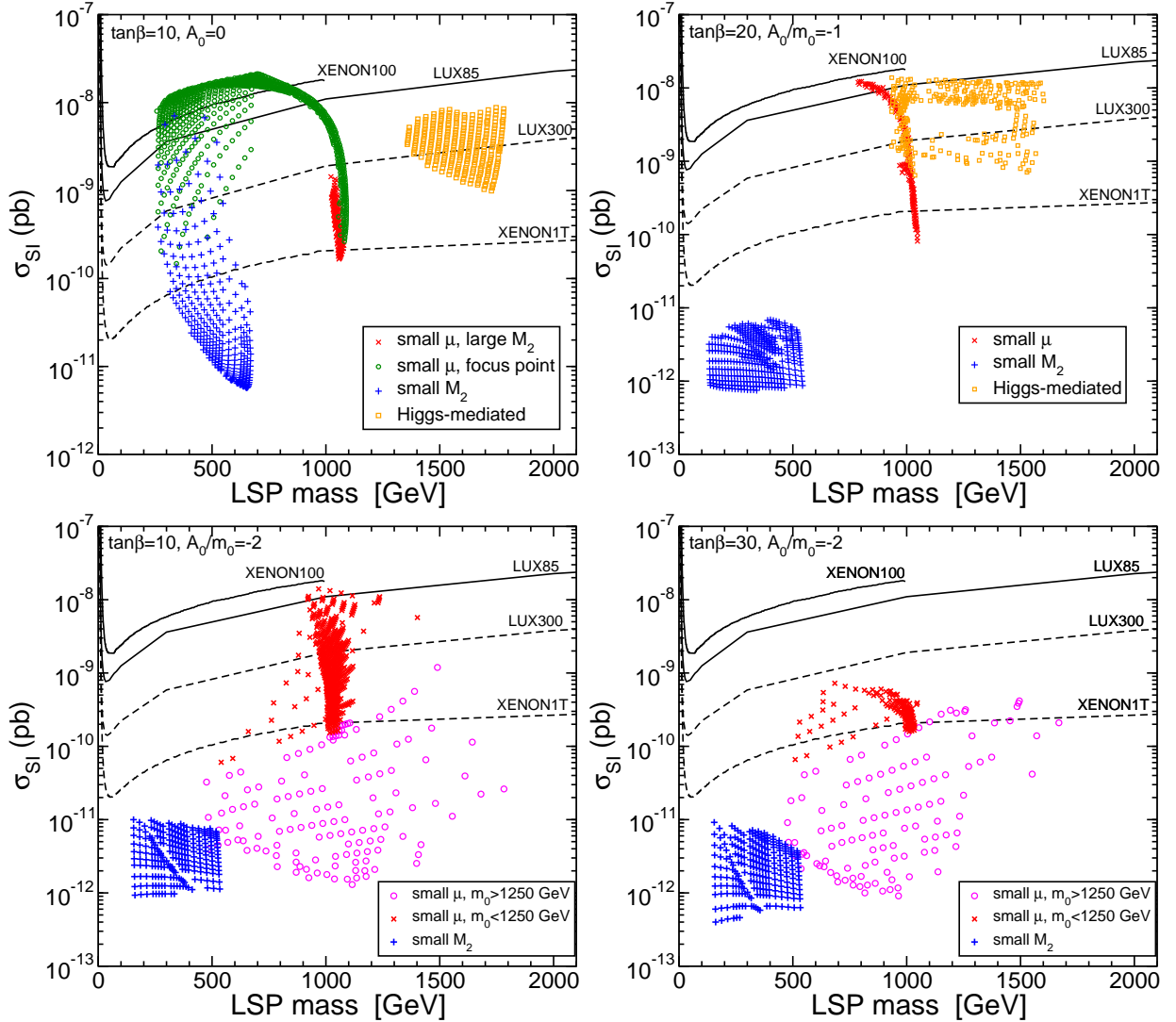

 \begin{minipage}[]{0.49\linewidth}
   \includegraphics[width=\linewidth,angle=0]{sigmaSI_0_10.eps}
 \end{minipage}
 \begin{minipage}[]{0.49\linewidth}
   \includegraphics[width=\linewidth,angle=0]{sigmaSI_-1_20.eps}
 \end{minipage}
 \begin{minipage}[]{0.49\linewidth}
   \includegraphics[width=\linewidth,angle=0]{sigmaSI_-2_10.eps}
 \end{minipage}
 \begin{minipage}[]{0.49\linewidth}
   \includegraphics[width=\linewidth,angle=0]{sigmaSI_-2_30.eps}
 \end{minipage}
\caption{\label{fig:sigmaSI}
The spin-independent LSP-nucleon cross-section as a function of LSP mass, for models shown in Figures 
\ref{fig:tb10_Arat0}, \ref{fig:tb20_Arat-1}, \ref{fig:tb30_Arat-2}, and \ref{fig:tb10_Arat-2} that satisfy
$\Omega_{\rm DM} h^2=0.120\pm 0.005$.
The model point symbols are coded according to
four regions: the small $\mu$ region with large $M_2$ at positive $\theta_{24}$ (red); the focus point region
with small $\mu$ and $m_0> 5000$ GeV (green), which occurs only in the first panel; the small $M_2$ region with negative $\theta_{24}$ (blue); 
and the Higgs-mediated co-annihilation island region (orange), which occurs only in the
top two panels. In the bottom two panels, the small
$\mu$ region points with $m_0$ greater than 1250 GeV are shown separately (pink circles).} 
\end{figure}

Finally, I turn to the dark-matter motivated alternative in which one requires 
a thermal relic abundance in accord with the WMAP and Planck 
observations. This requires a larger value of $|\mu|/M_1$ in order to 
avoid overly efficient annihilation of the dark matter, so the 
superpartner mass spectrum looks qualitatively similar to Figure 
\ref{fig:spectrum}, except that the higgsino-like states will be much 
heavier, of order 1000 GeV. I show in the first panel of Figure 
\ref{fig:sigmaSI} the predictions for the spin-independent LSP-nucleon 
cross-sections (obtained from version 3.1 of micrOmegas 
\cite{micromegas1}-\cite{micromegas4} with the default choices for 
nuclear matrix elements, including $f_s = 0.0447$) 
for the models of Figure \ref{fig:tb10_Arat0} 
with $\tan\beta = 10$, varying $\theta_{24}$ and $m_0$, and $M_3$ 
continuously varied from $1200$ to $2500$ GeV. Here, by varying 
$\theta_{24}$, I require that the thermal relic abundance lies in the 
range $\Omega_{\rm DM} h^2 = 0.120 \pm 0.005$. The model point symbols are coded 
according to four regions: the small $\mu$ region with large $M_2$ and 
$\theta_{24}/\pi > 0.05$ (red); the focus point region with small $\mu$ and 
$m_0> 5000$ GeV (green); the small $M_2$ region with $\theta_{24}/\pi \lsim -0.055$ (blue); and 
the Higgs-mediated co-annihilation island region (orange). The boundaries between 
the first and second regions and the second and third regions are fuzzy, 
as indicated by the overlap of model points. Also shown are the present 
limits from XENON100 \cite{XENON100} and the LUX 85 day data \cite{LUX} 
(solid lines), and some projected reaches for LUX 300 day \cite{LUX300d} 
and XENON 1T \cite{XENON1T} runs (dashed lines). Some, but not all, of the 
focus point models are in tension with the XENON100 and LUX85 limits. 
(Here it is good to keep in mind the significant uncertainties associated 
with nuclear matrix elements.) However, the other regions are clearly 
safely beyond the current limits. Most of the small $\mu$ region and 
the Higgs-mediated island will eventually be explored by ton-class direct 
detection experiments, but the bino-wino co-annihilation region will 
continue to be a challenge.

Similarly, the remaining three panels of Figure \ref{fig:sigmaSI} show the corresponding spin-independent
LSP-nucleon cross-sections for the other three cases that were studied above in 
Figures \ref{fig:tb20_Arat-1}, \ref{fig:tb30_Arat-2}, and \ref{fig:tb10_Arat-2}. These cases do not have 
focus point regions, but are otherwise qualitatively similar to the first panel. 
In the cases of $A_0 = -2m_0$, the regions with small $\mu$ that have $m_0 > 1250$ GeV 
are indicated separately, and mostly consist of models where the dark matter thermal 
abundance arises in part due to co-annihilations
with top squarks. We see that the region with
relatively small $\mu$ will be nearly, but not completely, probed by ton-class direct dark matter detection experiments, provided that $m_0$ is not larger than roughly 1250 GeV, depending on the other parameters
of the model. For the larger $m_0$ case with stop co-annihilation, and for the wino-coannihilation case, 
it will remain very challenging for direct detection experiments to probe the models for some time. 

\section{Outlook\label{sec:outlook}}
\setcounter{equation}{0}
\setcounter{figure}{0}
\setcounter{table}{0}
\setcounter{footnote}{1}

The discovery of the Higgs boson with a mass near 126 GeV is consistent 
with supersymmetry, and provides an important clue as to the superpartner 
mass spectrum.\footnote{It is interesting that for any given model 
framework, the usefulness of this clue is already limited by the 
theoretical uncertainties in the Higgs mass computation, which one can 
hope will be reduced with further work.} This clue fits nicely with an 
expectation that the top squarks should be heavy. A heavy spectrum of 
squark and slepton masses also fits well with the current non-observation of
squarks and the gluino at the LHC, and the suppression of flavor 
violation that one might otherwise expect in supersymmetry. However, it is in 
tension with expectations from naturalness. In this paper I have embraced
the idea of only requiring ``semi-natural" supersymmetry, where $|\mu|$
is required to be of order a few hundred GeV or less, but no expectations
are taken with regard to other superpartner masses. 

Within the framework of non-universal gaugino masses at the scale of gauge 
coupling unification, this is seen to be nicely consistent with the idea 
of flavor-preserving sfermion masses coming from gaugino mass dominance, 
$m_{1/2}^2 \gg m_0^2, A_0^2$, like the well-known ``no-scale" or ``gaugino 
mediation" ideas. Although this idea was only mapped out in this paper for 
a tiny subset of possible models, it should be clear that this will work 
much more generally. For example, one can see from Figure 
\ref{fig:tb10_Arat0_2000ALL} that the same thing can occur for 
$\theta_{24}/\pi \approx 0.33$, a region of parameter space that is far 
from the CMSSM case. The region of parameter space that we may be lead 
to through the clues mentioned above can be quite challenging both for the 
future explorations of the LHC and for direct dark matter experiments.

{\it Acknowledgments:} I thank Graham Kribs and James Younkin for relevant conversations. 
This work was supported in part by the National Science Foundation grant 
number PHY-1068369. This research was supported in part by the National 
Science Foundation under Grant No. NSF PHY11-25915.



\begin{thebibliography}{90}
\baselineskip=13.925pt

\bibitem{ATLASSUSY1} 
ATLAS collaboration,
``Search for squarks and gluinos with the ATLAS detector 
in final states with jets and missing transverse momentum 
and 20.3 fb$^{-1}$ of $\sqrt{s}=8$ TeV proton-proton collision data,"
ATLAS-CONF-2013-047, May 2013.

\bibitem{ATLASSUSY2} 
ATLAS collaboration,
``Search for squarks and gluinos in events with isolated leptons, 
jets and missing transverse momentum at $\sqrt{s}=8$ 
TeV with the ATLAS detector,"
ATLAS-CONF-2013-062, June 2013;

\bibitem{ATLASSUSY3} 
G.~Aad {\it et al.}  [ ATLAS Collaboration],
``Search for new phenomena in final states with large jet 
multiplicities and 
missing transverse momentum at sqrt(s)=8 TeV 
proton-proton collisions using the ATLAS experiment,''
[1308.1841].

\bibitem{CMSSUSY}
CMS Collaboration,
``Search for New Physics in the Multijets and 
Missing Momentum Final State in Proton-Proton 
Collisions at 8 TeV," CMS-PAS-SUS-13-012.

\bibitem{Alves1} 
  D.~S.~M.~Alves, E.~Izaguirre and J.~G.~Wacker,
  Phys.\ Lett.\ B {\bf 702}, 64 (2011)
  [1008.0407].
  
\bibitem{Alves2}  
  D.~S.~M.~Alves, E.~Izaguirre and J.~G.~Wacker,
  JHEP {\bf 1110}, 012 (2011)
  [1102.5338].
    
\bibitem{LeCompte1} 
  T.~J.~LeCompte and S.~P.~Martin,
  Phys.\ Rev.\ D {\bf 84}, 015004 (2011)
  [1105.4304 [hep-ph]].
  
\bibitem{LeCompte2}  
  T.~J.~LeCompte and S.~P.~Martin,
  Phys.\ Rev.\ D {\bf 85}, 035023 (2012)
  [1111.6897 [hep-ph]].

\bibitem{Baer:2012up} 
  H.~Baer, V.~Barger, P.~Huang, A.~Mustafayev and X.~Tata,
  Phys.\ Rev.\ Lett.\  {\bf 109}, 161802 (2012)
  [1207.3343 [hep-ph]].
  
\bibitem{Feng:2013pwa} 
  J.~L.~Feng,
  ``Naturalness and the Status of Supersymmetry,''
  [1302.6587].

\bibitem{Craig:2013cxa} 
  N.~Craig,
  ``The State of Supersymmetry after Run I of the LHC,''
  1309.0528 [hep-ph].

\bibitem{Baer:2013gva} 
  H.~Baer, V.~Barger and D.~Mickelson,
  ``How conventional measures overestimate electroweak fine-tuning in supersymmetric theory,''
  Phys.\ Rev.\ D {\bf 88}, 095013 (2013)
  [1309.2984 [hep-ph]].
  
\bibitem{KimNilles}
J.E.~Kim and H.~P.~Nilles,
  Phys.\ Lett.\ B {\bf 138}, 150 (1984).

\bibitem{GiudiceMasiero}
G.F.~Giudice and A.~Masiero,
  Phys.\ Lett.\ B {\bf 206}, 480 (1988).

\bibitem{NMSSMreview} 
  U.~Ellwanger, C.~Hugonie and A.~M.~Teixeira,
  Phys.\ Rept.\  {\bf 496}, 1 (2010)
  [0910.1785 [hep-ph]].

\bibitem{Papucci:2011wy} 
  M.~Papucci, J.~T.~Ruderman and A.~Weiler,
  JHEP {\bf 1209}, 035 (2012)
  [1110.6926 [hep-ph]].

\bibitem{Evans:2013jna} 
  J.~A.~Evans, Y.~Kats, D.~Shih and M.~J.~Strassler,
  ``Toward Full LHC Coverage of Natural Supersymmetry,''
  1310.5758 [hep-ph].

\bibitem{higgsinoworld1}
  G.~L.~Kane and J.~D.~Wells,
  Phys.\ Rev.\ Lett.\  {\bf 76}, 4458 (1996)
  [hep-ph/9603336].

\bibitem{higgsinoworld2}
G. L. Kane, in {\em Kane, G.L. (ed.): Perspectives on supersymmetry} 352-354;  
  
\bibitem{Baer:2011ec} 
  H.~Baer, V.~Barger and P.~Huang,
  ``Hidden SUSY at the LHC: the light higgsino-world scenario and the role of a lepton collider,''
  JHEP {\bf 1111}, 031 (2011)
  [1107.5581 [hep-ph]].

\bibitem{Han:2013usa} 
  C.~Han, A.~Kobakhidze, N.~Liu, A.~Saavedra, L.~Wu and J.~M.~Yang,
  ``Probing Light Higgsinos in Natural SUSY from Monojet Signals at the LHC,''
  1310.4274 [hep-ph].

\bibitem{focuspoint1}
  K.~L.~Chan, U.~Chattopadhyay and P.~Nath,
  Phys.\ Rev.\ D {\bf 58}, 096004 (1998)
  [hep-ph/9710473],

\bibitem{focuspoint2}
J.~L.~Feng, K.~T.~Matchev and T.~Moroi,
  Phys.\ Rev.\ Lett.\  {\bf 84}, 2322 (2000)
  [hep-ph/9908309];
  
\bibitem{focuspoint3}
J.~L.~Feng, K.~T.~Matchev and T.~Moroi,
  Phys.\ Rev.\ D {\bf 61}, 075005 (2000)
  [hep-ph/9909334].

\bibitem{focuspoint4}
J.~L.~Feng, K.~T.~Matchev and F.~Wilczek,
  Phys.\ Lett.\ B {\bf 482}, 388 (2000)
  [hep-ph/0004043].

\bibitem{focuspoint5} 
  J.~L.~Feng, K.~T.~Matchev and D.~Sanford,
  Phys.\ Rev.\ D {\bf 85}, 075007 (2012)
  [1112.3021].
    
\bibitem{focuspoint6} 
  J.~L.~Feng and D.~Sanford,
  Phys.\ Rev.\ D {\bf 86}, 055015 (2012)
  [1205.2372].

\bibitem{KaneKing1}
  G.L.~Kane and S.F.~King,
  Phys.\ Lett.\  B {\bf 451}, 113 (1999)
  [hep-ph/9810374],
  
\bibitem{KaneKing2}  
  M.~Bastero-Gil, G.L.~Kane and S.F.~King,
  Phys.\ Lett.\  B {\bf 474}, 103 (2000)
  [hep-ph/9910506].

\bibitem{Huitu:2005wh} 
  K.~Huitu, J.~Laamanen, P.~N.~Pandita and S.~Roy,
  Phys.\ Rev.\ D {\bf 72}, 055013 (2005)
  [hep-ph/0502100].
  
\bibitem{Baer:2006dz} 
  H.~Baer, A.~Mustafayev, E.~-K.~Park, S.~Profumo and X.~Tata,
  JHEP {\bf 0604}, 041 (2006)
  [hep-ph/0603197].

\bibitem{Huitu:2006sg} 
  K.~Huitu, J.~Laamanen, P.~N.~Pandita and S.~Roy,
  Pramana {\bf 69}, 819 (2007)
  [hep-ph/0607311].
  
\bibitem{Abe:2007kf} 
  H.~Abe, T.~Kobayashi and Y.~Omura,
  Phys.\ Rev.\ D {\bf 76}, 015002 (2007)
  [hep-ph/0703044 [HEP-PH]].

\bibitem{compressedSUSY}
S.P.~Martin,
  Phys.\ Rev.\  D {\bf 75}, 115005 (2007)
  [hep-ph/0703097],

\bibitem{Ananthanarayan:2007fj} 
  B.~Ananthanarayan and P.~N.~Pandita,
  Int.\ J.\ Mod.\ Phys.\ A {\bf 22}, 3229 (2007)
  [0706.2560 [hep-ph]].
  
\bibitem{compressedSUSY2}
S.P.~Martin,
  Phys.\ Rev.\  D {\bf 76}, 095005 (2007)
  [hep-ph/0707.2812],

\bibitem{nonudma}
  S.F.~King, J.P.~Roberts and D.~P.~Roy,
  JHEP\ {\bf 0710}, 106  (2007)
  [0705.4219 [hep-ph]].
  
\bibitem{BMST}
H.~Baer, A.~Mustafayev, H.~Summy and X.~Tata,
  JHEP\ {\bf 0710}, 088  (2007)
  [0708.4003 [hep-ph]].
  
\bibitem{Horton:2009ed} 
  D.~Horton and G.~G.~Ross,
  Nucl.\ Phys.\ B {\bf 830}, 221 (2010)
  [0908.0857].

\bibitem{Okada:2011wd} 
  N.~Okada, S.~Raza and Q.~Shafi,
  Phys.\ Rev.\ D {\bf 84}, 095018 (2011)
  [1107.0941].

\bibitem{Younkin:2012ui} 
  J.~E.~Younkin and S.~P.~Martin,
  Phys.\ Rev.\ D {\bf 85}, 055028 (2012)
  [1201.2989].

\bibitem{Brummer:2012zc} 
  F.~Brummer and W.~Buchmuller,
  JHEP {\bf 1205}, 006 (2012)
  [1201.4338 [hep-ph]].
    
\bibitem{Caron:2012sf} 
  S.~Caron, J.~Laamanen, I.~Niessen and A.~Strubig,
  JHEP {\bf 1206}, 008 (2012)
  [1202.5288].

\bibitem{Antusch:2012gv} 
  S.~Antusch, L.~Calibbi, V.~Maurer, M.~Monaco and M.~Spinrath,
  JHEP {\bf 01}, 187 (2013)
  [1207.7236].

\bibitem{Gogoladze:2012yf} 
  I.~Gogoladze, F.~Nasir and Q.~Shafi,
  Int.\ J.\ Mod.\ Phys.\ A {\bf 28}, 1350046 (2013)
  [1212.2593 [hep-ph]].
  
\bibitem{Spies:2013fba} 
  A.~Spies and G.~Anton,
  JCAP {\bf 1306}, 022 (2013)
  [1306.1099]. 
  
\bibitem{Gogoladze:2013wva} 
  I.~Gogoladze, F.~Nasir and Q.~Shafi,
  ``SO(10) as a Framework for Natural Supersymmetry,''
  [1306.5699].

\bibitem{Yanagida:2013ah} 
  T.~T.~Yanagida and N.~Yokozaki,
  Phys.\ Lett.\ B {\bf 722}, 355 (2013)
  [1301.1137].

\bibitem{Miller:2013jra} 
  D.~J.~Miller and A.~P.~Morais,
  ``Supersymmetric SU(5) Grand Unification for a Post Higgs Boson Era,''
  [1307.1373].

\bibitem{Ajaib:2013kka} 
  M.~A.~Ajaib, I.~Gogoladze and Q.~Shafi,
  ``Sparticle Spectroscopy from SO(10) GUT with a Unified Higgs Sector,''
  [1307.4882].

\bibitem{Badziak:2013eda} 
  M.~Badziak, M.~Olechowski and S.~Pokorski,
  ``Light staus and enhanced Higgs diphoton rate with non-universal gaugino masses and SO(10) Yukawa unification,''
  [1307.7999].
      
\bibitem{Yanagida:2013uka} 
  T.~T.~Yanagida and N.~Yokozaki,
  ``Bino-Higgsino Mixed Dark Matter in a Focus Point Gaugino Mediation,''
  [1308.0536].

\bibitem{Kaminska:2013mya} 
  A.~Kaminska, G.~G.~Ross and K.~Schmidt-Hoberg,
  ``Non-universal gaugino masses and fine tuning implications for SUSY searches in the MSSM and the GNMSSM,''
  [1308.4168].

\bibitem{Cabrera:2013jya} 
  M.~E.~Cabrera, A.~Casas, R.~R.~de Austri and G.~Bertone,
  ``LHC and dark matter phenomenology of the NUGHM,''
  1311.7152 [hep-ph].

\bibitem{SU5nonuniversal1}
J.R.~Ellis, C.~Kounnas and D.V.~Nanopoulos,
  Nucl.\ Phys.\  B {\bf 247}, 373 (1984),
  
\bibitem{SU5nonuniversal2}  
J.R.~Ellis, K.~Enqvist, D.V.~Nanopoulos and K.~Tamvakis,
  Phys.\ Lett.\  B {\bf 155}, 381 (1985),

\bibitem{SU5nonuniversal3}
M.~Drees,
  Phys.\ Lett.\  B {\bf 158}, 409 (1985),

\bibitem{SU5nonuniversal4}
G.~Anderson, C.H.~Chen, J.F.~Gunion, J.D.~Lykken, T.~Moroi and Y.~Yamada,
  ``Motivations for and implications of non-universal GUT-scale boundary
  conditions for soft SUSY-breaking parameters,''
{\it In the Proceedings of 1996 DPF / DPB Summer Study on New Directions
for High-Energy Physics (Snowmass 96)},
  [hep-ph/9609457].
  
\bibitem{SU5nonuniversal5}  
G.~Anderson, H.~Baer, C.h.~Chen and X.~Tata,
  Phys.\ Rev.\  D {\bf 61}, 095005 (2000)
  [hep-ph/9903370].

\bibitem{SO10nonuniversal1}
J.~Chakrabortty and A.~Raychaudhuri,
  Phys.\ Lett.\  B {\bf 673}, 57 (2009)
  [hep-ph/0812.2783],

\bibitem{SO10nonuniversal2}
S.P.~Martin,
  Phys.\ Rev.\  D {\bf 79}, 095019 (2009)
  [hep-ph/0903.3568].
  
\bibitem{ATLASHiggs}
G.~Aad {\it et al.}  [ATLAS Collaboration],
  ``Observation of a new particle in the search for the Standard Model Higgs boson with the ATLAS detector at the LHC,''
  [1207.7214],

\bibitem{CMSHiggs}
  S.~Chatrchyan {\it et al.}  [CMS Collaboration],
  ``Observation of a new boson at a mass of 125 GeV with the CMS experiment at the LHC,''
  Phys.\ Lett.\ B {\bf 716}, 30 (2012)
  [1207.7235].

\bibitem{ATLAScombination}
  [ATLAS Collaboration],
  ``Combined measurements of the mass and signal strength of the Higgs-like boson with the ATLAS detector using up to 25 fb$^{-1}$ of proton-proton collision data,''
  ATLAS-CONF-2013-014, March 6, 2013.
  
\bibitem{CMScombination} 
  [CMS Collaboration],
  ``Combination of standard model Higgs boson searches and measurements of the properties of the new boson with a mass near 125 GeV''
  CMS-PAS-HIG-12-045, November 16, 2012.

\bibitem{Ellis:1990nz} 
  J.~R.~Ellis, G.~Ridolfi and F.~Zwirner,
  Phys.\ Lett.\ B {\bf 257}, 83 (1991).

\bibitem{Okada:1990vk} 
  Y.~Okada, M.~Yamaguchi and T.~Yanagida,
  Prog.\ Theor.\ Phys.\  {\bf 85}, 1 (1991).

\bibitem{Haber:1990aw} 
  H.~E.~Haber and R.~Hempfling,
  Phys.\ Rev.\ Lett.\  {\bf 66}, 1815 (1991).

\bibitem{Brignole:1992uf} 
  A.~Brignole,
  Phys.\ Lett.\ B {\bf 281}, 284 (1992).

\bibitem{Chankowski:1992ej} 
  P.~H.~Chankowski, S.~Pokorski and J.~Rosiek,
  Phys.\ Lett.\ B {\bf 286}, 307 (1992).

\bibitem{Chankowski:1992er} 
  P.~H.~Chankowski, S.~Pokorski and J.~Rosiek,
  Nucl.\ Phys.\ B {\bf 423}, 437 (1994)
  [hep-ph/9303309].
        
\bibitem{Hempfling:1993qq} 
  R.~Hempfling and A.~H.~Hoang,
  Phys.\ Lett.\ B {\bf 331}, 99 (1994)
  [hep-ph/9401219].
  
\bibitem{Casas:1994us} 
  J.~A.~Casas, J.~R.~Espinosa, M.~Quiros and A.~Riotto,
  Nucl.\ Phys.\ B {\bf 436}, 3 (1995)
  [Erratum-ibid.\ B {\bf 439}, 466 (1995)]
  [hep-ph/9407389].

\bibitem{Dabelstein:1994hb} 
  A.~Dabelstein,
  Z.\ Phys.\ C {\bf 67}, 495 (1995)
  [hep-ph/9409375].
    
\bibitem{Carena:1995bx} 
  M.~S.~Carena, J.~R.~Espinosa, M.~Quiros and C.~E.~M.~Wagner,
  Phys.\ Lett.\ B {\bf 355}, 209 (1995)
  [hep-ph/9504316].
  
\bibitem{Carena:1995wu} 
  M.~S.~Carena, M.~Quiros and C.~E.~M.~Wagner,
  Nucl.\ Phys.\ B {\bf 461}, 407 (1996)
  [hep-ph/9508343].

\bibitem{Heinemeyer:1998jw} 
  S.~Heinemeyer, W.~Hollik and G.~Weiglein,
  Phys.\ Rev.\ D {\bf 58}, 091701 (1998)
  [hep-ph/9803277].

\bibitem{Heinemeyer:1998kz} 
  S.~Heinemeyer, W.~Hollik and G.~Weiglein,
  Phys.\ Lett.\ B {\bf 440}, 296 (1998)
  [hep-ph/9807423].

\bibitem{Zhang:1998bm} 
  R.~-J.~Zhang,
  Phys.\ Lett.\ B {\bf 447}, 89 (1999)
  [hep-ph/9808299].

\bibitem{Espinosa:1999zm} 
  J.~R.~Espinosa and R.~-J.~Zhang,
  JHEP {\bf 0003}, 026 (2000)
  [hep-ph/9912236].
     
\bibitem{Carena:2000dp} 
  M.~S.~Carena, H.~E.~Haber, S.~Heinemeyer, W.~Hollik, C.~E.~M.~Wagner and G.~Weiglein,
  Nucl.\ Phys.\ B {\bf 580}, 29 (2000)
  [hep-ph/0001002].

\bibitem{Espinosa:2000df} 
  J.~R.~Espinosa and R.~-J.~Zhang,
  Nucl.\ Phys.\ B {\bf 586}, 3 (2000)
  [hep-ph/0003246].

\bibitem{Degrassi:2001yf} 
  G.~Degrassi, P.~Slavich and F.~Zwirner,
  Nucl.\ Phys.\ B {\bf 611}, 403 (2001)
  [hep-ph/0105096].
  
\bibitem{Brignole:2001jy} 
  A.~Brignole, G.~Degrassi, P.~Slavich and F.~Zwirner,
  Nucl.\ Phys.\ B {\bf 631}, 195 (2002)
  [hep-ph/0112177].
  
\bibitem{Brignole:2002bz} 
  A.~Brignole, G.~Degrassi, P.~Slavich and F.~Zwirner,
  Nucl.\ Phys.\ B {\bf 643}, 79 (2002)
  [hep-ph/0206101].

\bibitem{Martin:2001vx} 
  S.~P.~Martin,
  Phys.\ Rev.\ D {\bf 65}, 116003 (2002)
  [hep-ph/0111209].

\bibitem{Martin:2002iu} 
  S.~P.~Martin,
  Phys.\ Rev.\ D {\bf 66}, 096001 (2002)
  [hep-ph/0206136].
    
\bibitem{Martin1} 
  S.~P.~Martin,
  Phys.\ Rev.\ D {\bf 67}, 095012 (2003)
  [hep-ph/0211366].
  
\bibitem{Martin:2003qz} 
  S.~P.~Martin,
  Phys.\ Rev.\ D {\bf 68}, 075002 (2003)
  [hep-ph/0307101].

\bibitem{Martin2}  
  S.~P.~Martin,
  Phys.\ Rev.\ D {\bf 71}, 016012 (2005)
  [hep-ph/0405022].
  
\bibitem{Martin:2005qm} 
  S.~P.~Martin and D.~G.~Robertson,
  Comput.\ Phys.\ Commun.\  {\bf 174}, 133 (2006)
  [hep-ph/0501132].
  
\bibitem{Martin:2007pg} 
  ``Three-loop corrections to the lightest Higgs scalar boson mass in supersymmetry,''
  Phys.\ Rev.\ D {\bf 75}, 055005 (2007)
  [hep-ph/0701051].

\bibitem{Harlander:2008ju} 
  R.~V.~Harlander, P.~Kant, L.~Mihaila and M.~Steinhauser,
  Phys.\ Rev.\ Lett.\  {\bf 100}, 191602 (2008)
  [Phys.\ Rev.\ Lett.\  {\bf 101}, 039901 (2008)]
  [0803.0672 [hep-ph]].

\bibitem{feynhiggs1} 
  S.~Heinemeyer, W.~Hollik and G.~Weiglein,
  Comput.\ Phys.\ Commun.\  {\bf 124}, 76 (2000)
  [hep-ph/9812320].

\bibitem{feynhiggs2}   
  S.~Heinemeyer, W.~Hollik and G.~Weiglein,
  Eur.\ Phys.\ J.\ C {\bf 9}, 343 (1999)
  [hep-ph/9812472].

\bibitem{feynhiggs3}    
G.~Degrassi, S.~Heinemeyer, W.~Hollik, P.~Slavich and G.~Weiglein,
  Eur.\ Phys.\ J.\ C {\bf 28}, 133 (2003)
  [hep-ph/0212020].

\bibitem{feynhiggs4}   
M.~Frank, T.~Hahn, S.~Heinemeyer, W.~Hollik, H.~Rzehak and G.~Weiglein,
  JHEP {\bf 0702}, 047 (2007)
  [hep-ph/0611326].

\bibitem{feynhiggs5}
 T.~Hahn, S.~Heinemeyer, W.~Hollik, H.~Rzehak and G.~Weiglein,
  Comput.\ Phys.\ Commun.\  {\bf 180}, 1426 (2009).
         
\bibitem{softsusy} 
  B.~C.~Allanach,
  Comput.\ Phys.\ Commun.\  {\bf 143}, 305 (2002)
  [hep-ph/0104145].

\bibitem{suspect}
  A.~Djouadi, J.L.~Kneur and G.~Moultaka,
  Comput.\ Phys.\ Commun.\  {\bf 176}, 426 (2007)
  [hep-ph/0211331].

\bibitem{spheno} 
  W.~Porod,
  Comput.\ Phys.\ Commun.\  {\bf 153}, 275 (2003)
  [hep-ph/0301101].

\bibitem{CPsuperH1} 
  J.~S.~Lee, A.~Pilaftsis, M.~S.~Carena, S.~Y.~Choi, M.~Drees, J.~R.~Ellis and C.~E.~M.~Wagner,
  Comput.\ Phys.\ Commun.\  {\bf 156}, 283 (2004)
  [hep-ph/0307377].
  
\bibitem{CPsuperH2}
  J.~S.~Lee, M.~Carena, J.~Ellis, A.~Pilaftsis and C.~E.~M.~Wagner,
  Comput.\ Phys.\ Commun.\  {\bf 180} (2009) 312
  [0712.2360 [hep-ph]].

\bibitem{CPsuperH3} 
  J.~S.~Lee, M.~Carena, J.~Ellis, A.~Pilaftsis and C.~E.~M.~Wagner,
  Comput.\ Phys.\ Commun.\  {\bf 184}, 1220 (2013)
  [1208.2212 [hep-ph]].
  
\bibitem{Paige:2003mg} 
  F.~E.~Paige, S.~D.~Protopopescu, H.~Baer and X.~Tata,
  ``ISAJET 7.69: A Monte Carlo event generator for pp, anti-p p, and e+e- reactions,''
  hep-ph/0312045.

\bibitem{Kant:2010tf} 
  P.~Kant, R.~V.~Harlander, L.~Mihaila and M.~Steinhauser,
  ``Light MSSM Higgs boson mass to three-loop accuracy,''
  JHEP {\bf 1008}, 104 (2010)
  [1005.5709].

\bibitem{Feng:2013tvd} 
  J.~L.~Feng, P.~Kant, S.~Profumo and D.~Sanford,
  ``3-Loop Corrections to the Higgs Boson Mass and Implications for Supersymmetry at the LHC,''
  [1306.2318].
      
\bibitem{pmssm1}
C.~F.~Berger, J.~S.~Gainer, J.~L.~Hewett and T.~G.~Rizzo,
  JHEP {\bf 0902}, 023 (2009)
  [0812.0980 [hep-ph]].

\bibitem{pmssm2}
M.~W.~Cahill-Rowley, J.~L.~Hewett, A.~Ismail and T.~G.~Rizzo,
  Phys.\ Rev.\ D {\bf 86}, 075015 (2012)
  [1206.5800 [hep-ph]].
  
\bibitem{pmssm3} 
  M.~Cahill-Rowley, J.~L.~Hewett, A.~Ismail and T.~G.~Rizzo,
  ``pMSSM Studies at the 7, 8 and 14 TeV LHC,''
  1307.8444 [hep-ph].

\bibitem{micromegas1}
  G.~Belanger, F.~Boudjema, A.~Pukhov and A.~Semenov,
  Comput.\ Phys.\ Commun.\  {\bf 180}, 747 (2009)
  [hep-ph/0803.2360],
  
\bibitem{micromegas2}
  G.~Belanger, F.~Boudjema, A.~Pukhov and A.~Semenov,  
  Comput.\ Phys.\ Commun.\  {\bf 176}, 367 (2007)
  [hep-ph/0607059],

\bibitem{micromegas3}
  G.~Belanger, F.~Boudjema, A.~Pukhov and A.~Semenov,
  Comput.\ Phys.\ Commun.\  {\bf 174}, 577 (2006)
  [hep-ph/0405253],

\bibitem{micromegas4}
  G.~Belanger, F.~Boudjema, A.~Pukhov and A.~Semenov,
  Comput.\ Phys.\ Commun.\  {\bf 149}, 103 (2002)
  [hep-ph/0112278].

\bibitem{Covi:1999ty}
  L.~Covi, J.~E.~Kim and L.~Roszkowski,
  Phys.\ Rev.\ Lett.\  {\bf 82}, 4180 (1999)
  [hep-ph/9905212].

\bibitem{Covi:2001nw}
  L.~Covi, H.~-B.~Kim, J.~E.~Kim and L.~Roszkowski,
  JHEP {\bf 0105}, 033 (2001)
  [hep-ph/0101009].

\bibitem{Feng:2003xh}
  J.~L.~Feng, A.~Rajaraman and F.~Takayama,
  Phys.\ Rev.\ Lett.\  {\bf 91}, 011302 (2003)
  [hep-ph/0302215].

\bibitem{Choi:2011yf}
  K.~-Y.~Choi, L.~Covi, J.~E.~Kim and L.~Roszkowski,
  JHEP {\bf 1204}, 106 (2012)
  [1108.2282 [hep-ph]].

\bibitem{XENON100} 
  E.~Aprile {\it et al.}  [XENON100 Collaboration],
  ``Dark Matter Results from 225 Live Days of XENON100 Data,''
  Phys.\ Rev.\ Lett.\  {\bf 109}, 181301 (2012)
  [1207.5988].

\bibitem{LUX}
  D.~S.~Akerib {\it et al.}  [LUX Collaboration],
  ``First results from the LUX dark matter experiment at the Sanford Underground Research Facility,''
  [1310.8214].

\bibitem{WMAP}
C.~L.~Bennett {\it et al.}  [WMAP Collaboration],
  ``Nine-Year Wilkinson Microwave Anisotropy Probe (WMAP) Observations: Final Maps and Results,''
  Astrophys.\ J.\ Suppl.\  {\bf 208}, 20 (2013)
  [1212.5225].

\bibitem{Planck}
  P.~A.~R.~Ade {\it et al.}  [Planck Collaboration],
  ``Planck 2013 results. XVI. Cosmological parameters,''
  [1303.5076].

\bibitem{noscale1}
  E.~Cremmer, S.~Ferrara, C.~Kounnas and D.~V.~Nanopoulos,
  Phys.\ Lett.\ B {\bf 133}, 61 (1983).

\bibitem{noscale2} 
  J.~R.~Ellis, C.~Kounnas and D.~V.~Nanopoulos,
  Nucl.\ Phys.\ B {\bf 247}, 373 (1984).

\bibitem{noscale3} 
  A.~B.~Lahanas and D.~V.~Nanopoulos,
  Phys.\ Rept.\  {\bf 145}, 1 (1987).

\bibitem{gauginomediation1} 
K.~Inoue, M.~Kawasaki, M.~Yamaguchi and T.~Yanagida,
  Phys.\ Rev.\ D {\bf 45}, 328 (1992).

\bibitem{gauginomediation2} 
D.~E.~Kaplan, G.~D.~Kribs and M.~Schmaltz,
  Phys.\ Rev.\ D {\bf 62}, 035010 (2000)
  [hep-ph/9911293].

\bibitem{gauginomediation3} 
Z.~Chacko, M.~A.~Luty, A.~E.~Nelson and E.~Ponton,
  JHEP {\bf 0001}, 003 (2000)
  [hep-ph/9911323].

\bibitem{winocoannihilation}
A.~Birkedal-Hansen and B.D.~Nelson,
  Phys.\ Rev.\ D {\bf 64}, 015008 (2001)
  [hep-ph/0102075].

\bibitem{Baer:2005zc} 
  H.~Baer, A.~Mustafayev, E.~-K.~Park and S.~Profumo,
  JHEP {\bf 0507}, 046 (2005)
  [hep-ph/0505227].

\bibitem{Baer:2005jq} 
  H.~Baer, T.~Krupovnickas, A.~Mustafayev, E.~-K.~Park, S.~Profumo and X.~Tata,
  JHEP {\bf 0512}, 011 (2005)
  [hep-ph/0511034].

\bibitem{Ibe:2013pua} 
  M.~Ibe, A.~Kamada and S.~Matsumoto,
  ``Mixed (Cold+Warm) Dark Matter in the Bino-Wino co-annihilation scenario,''
  1311.2162 [hep-ph].

\bibitem{Cheung:2012qy} 
  C.~Cheung, L.~J.~Hall, D.~Pinner and J.~T.~Ruderman,
  JHEP {\bf 1305}, 100 (2013)
  [1211.4873 [hep-ph]].
    
\bibitem{ArkaniHamed:2006mb} 
  N.~Arkani-Hamed, A.~Delgado and G.~F.~Giudice,
  Nucl.\ Phys.\ B {\bf 741}, 108 (2006)
  [hep-ph/0601041].

\bibitem{Drees}
M.~Drees and M.M.~Nojiri,
  Phys.\ Rev.\ D {\bf 47}, 376 (1993)
  [hep-ph/9207234].

\bibitem{stopco1} M.E.~Gomez, G.~Lazarides and C.~Pallis,
  Phys.\ Rev.\ D {\bf 61}, 123512 (2000)   
  [hep-ph/9907261];

\bibitem{stopco2} C.~Boehm, A.~Djouadi and M.~Drees,
  Phys.\ Rev.\ D {\bf 62}, 035012 (2000)
  [hep-ph/9911496];

\bibitem{stopco3} J.R.~Ellis, K.A.~Olive and Y.~Santoso,
  Astropart.\ Phys.\  {\bf 18}, 395 (2003)
  [hep-ph/0112113];

\bibitem{stopco4} C.~Balazs, M.~Carena and C.E.M.~Wagner,
  Phys.\ Rev.\ D {\bf 70}, 015007 (2004)
  [hep-ph/0403224].
 
\bibitem{Gori:2013ala} 
  S.~Gori, S.~Jung and L.~-T.~Wang,
  ``Cornering electroweakinos at the LHC,''
  1307.5952 [hep-ph].

\bibitem{Baer:2013yha} 
  H.~Baer, V.~Barger, P.~Huang, D.~Mickelson, A.~Mustafayev, W.~Sreethawong and X.~Tata,
  Phys.\ Rev.\ Lett.\  {\bf 110}, 151801 (2013)
  [1302.5816 [hep-ph]].

\bibitem{Han:2013kza} 
  T.~Han, S.~Padhi and S.~Su,
  ``Electroweakinos in the Light of the Higgs Boson,''
  1309.5966 [hep-ph].

\bibitem{WBF1} 
G.~F.~Giudice, T.~Han, K.~Wang and L.~-T.~Wang,
  Phys.\ Rev.\ D {\bf 81}, 115011 (2010)
  [1004.4902 [hep-ph]].

\bibitem{WBF2} 
B.~Dutta, A.~Gurrola, W.~Johns, T.~Kamon, P.~Sheldon and K.~Sinha,
  Phys.\ Rev.\ D {\bf 87}, 035029 (2013)
  [1210.0964 [hep-ph]].

\bibitem{WBF3} 
A.~G.~Delannoy, B.~Dutta, A.~Gurrola, W.~Johns, T.~Kamon, E.~Luiggi, A.~Melo and P.~Sheldon {\it et al.},
  Phys.\ Rev.\ Lett.\  {\bf 111}, 061801 (2013)
  [1304.7779 [hep-ph]].

\bibitem{LUX300d} D.~McKinsey and R.~Gaitskell, for the LUX Collaboration,
``First Science Results from the
LUX Dark Matter Experiment",
Sanford Underground Research Facility talk, October 30, 2013.

\bibitem{XENON1T} K.~Arisaka for the XENON Collaboration,
``The XENON Dark Matter Program at LNGS, XENON100, 1T, and beyond"
Snowmass on the Mississippi (CSS 2013) talk, March 7, 2013.

\end{thebibliography}
\end{document}